# Enhanced Magnetic-Inertial Odometry via Invariant Error-State Kalman Filtering with Scaled Magnetic-Field Coefficient-Bias Compensation

Chengdan Tan, Lubin Chang, Qiang Bian, and Yude Tong

***Abstract*—Magnetic-inertial odometry (MIO) exploits spatially varying magnetic anomalies to aid inertial navigation, enabling indoor navigation without a preconstructed magnetic-field map. Conventional MIO methods employ an error-state Kalman filter (ESKF) whose error-state equations are derived through first-order linearization around the current nominal estimate. Consequently, the linearized model may deviate from actual error propagation when this estimate is inaccurate. To address this issue, the attitude, velocity, and position are jointly formulated as elements of the Lie group $SE_2(3)$, and a right-invariant error is defined exactly on this group. In addition, a scaled magnetic-field polynomial-model coefficient bias is introduced to compensate for errors in propagating the polynomial-model coefficient. The strapdown inertial navigation system (SINS) error-state propagation and the magnetic-field model error-state propagation are rederived accordingly to construct a right-invariant error-state Kalman filter for MIO (RSE-KF-MIO). Experiments are conducted on public magnetometer-array datasets. Compared with the state-of-the-art open-source magnetic-field-aided inertial navigation system (MAINS), RSE-KF-MIO reduces the horizontal position, velocity, and attitude root-mean-square errors by approximately 28%–46% with position aiding, where short-term position measurements are provided by an additional positioning source, and by approximately 44%–63% without position aiding. Furthermore, sensor-count sweep experiments confirm the advantage of the proposed RSE-KF-MIO across different sensor configurations, and the ablation study further demonstrates the individual contributions of the right-invariant error modeling and the coefficient-bias compensation.**

## I. INTRODUCTION

RELIABLE indoor positioning remains challenging because global navigation satellite system (GNSS) signals are unavailable indoors. Among infrastructure-free solutions, strapdown inertial navigation system (SINS) is attractive because of its low cost and complete autonomy. However, the noise and bias of inertial sensors are accumulated through integration, causing the position error of a low-cost inertial measurement unit (IMU) to grow rapidly [1]. A common approach to addressing the challenge of the INS error accumulation is to fuse additional sensors, such as cameras, light detection and ranging (LiDAR), and wheel sensors. However, these approaches may suffer from degraded performance under specific conditions. For example, visual systems can achieve high accuracy, but their performance degrades considerably in low-light or weakly textured areas [2], [3]. LiDAR systems are insensitive to illumination changes, but their high sensor cost and computational overhead, together with their limited functionality in geometrically feature-poor areas, restrict their applicability [4]. Wheel navigation systems are inexpensive; however, sharp turns or low-adhesion surfaces can violate their nonholonomic constraints

This work was supported by the National Natural Science Foundation of China under Grant 62373367. (*Corresponding author: Lubin Chang*).
Chengdan Tan is with Naval University of Engineering, Wuhan 430033, China, also with Changsha Institute of Technology, Changsha 410200, China (e-mail: d26381105@nue.edu.cn).
Lubin Chang, Qiang Bian, Yude Tong are with Naval University of Engineering, Wuhan 430033, China (e-mail: 1410021052@nue.edu.cn; 0910021037@nue.edu.cn; 1310061107@nue.edu.cn).

and cause position drift [5]. A promising complementary approach is to exploit environment magnetic-field information acquired by magnetometers. Indoor facilities, such as reinforced-concrete structures, steel furniture, and electrical installations, distort the indoor geomagnetic field and produce spatially varying magnetic anomalies. Although such anomalies have been regarded as disturbances that degrade heading estimation in geomagnetic matching navigation, they also provide position-dependent information and thus serve as an effective source for indoor positioning [6], [7].

Indoor navigation techniques based on magnetic anomalies are roughly categorized into three types: magnetic fingerprinting methods, magnetic-field simultaneous localization and mapping (SLAM), and magnetic-field odometry. Magnetic fingerprinting methods [8], [9] generally require a preconstructed magnetic-field map, which limits their applicability and makes their performance dependent on map accuracy and resolution. Magnetic-field SLAM comprises two phases: exploration, in which a reference magnetic-field map is constructed, and exploitation, in which magnetic-field matching navigation is performed against this map. Previous studies have demonstrated meter-level indoor positioning accuracy using magnetic-field SLAM [10], [11], [12], [13]. However, these systems mostly rely on low-drift odometry during exploration, such as wheel or visual odometry. As discussed above, the performance of such odometry may degrade under specific conditions. If only a low-cost IMU is used to provide pose information during magnetic-field SLAM exploration, frequent loop closures are required, severely limiting exploration efficiency.

Magnetic-field odometry is a lightweight solution that provides odometric information using magnetometers. Since the local magnetic field is inhomogeneous over space, sensing this field supplies relative pose information without requiring a preconstructed map or additional infrastructure. In their seminal work, Vissière et al. [14] derived an equation relating body velocity to magnetic-field gradients and estimated body velocity from magnetometer-array measurements. Dorveaux et al. [15] extended this work by fusing IMU and distributed magnetometry measurements in a velocity estimator, thereby establishing a magnetic-inertial navigation framework. Early magnetic-field odometry methods relied on magnetic-field gradients to characterize the magnetic-field, instead of explicitly modeling the field itself. Wahlström et al. [16] originally proposed a Gaussian process (GP) method for global magnetic-field modeling. To reduce the computational cost of the GP model, Solin et al. [17] subsequently proposed a reduced-rank GP model. On the other hand, Skog et al. [18] developed a polynomial model to characterize the local magnetic-field under the curl-free and divergence-free assumptions, offering high computational efficiency. More recently, to capture local magnetic anomalies and complex nonlinear features, Liu et al. [19] proposed an equivalent magnetic dipole model and experimentally demonstrated its advantages.

The fusion of magnetic odometry with inertial measurements, referred to as magnetic-inertial odometry (MIO), has attracted considerable attention in indoor navigation. Chesneau et al. [20], [21] employed an extended Kalman filter (EKF) to improve magnetic-inertial odometry by incorporating inertial sensor biases and magnetic disturbances into the state vector. Zmitri et al. [22] proposed a magnetic-field-gradient-based EKF to improve velocity estimation by introducing a new differential equation for magnetic-field gradients to better capture magnetic-noise dynamics. Huang et al. [23] proposed the magnetic-field-aided inertial navigation system (MAINS), which establishes an ESKF-based tightly integrated framework and has become the state-of-the-art open-source MIO method. Building on this framework, Li et al. [24] and Liu et al. [19] improved MIO in terms of the filter structure and magnetic-field model, respectively, achieving improved positioning performance. However, these methods rely on the conventional ESKF framework: the nominal position, velocity, and attitude states propagate according to the nonlinear navigation equations, whereas the error states are obtained by first-order expansion around the current estimates. Consequently, an inaccurate nominal estimate directly compromises the accuracy of error-state propagation. Aggressive carrier dynamics, the use of a low-cost INS, and similar factors can render the nominal state estimate inaccurate, in which case the error-propagation model may deviate from the actual situation.

To mitigate the error-propagation inaccuracy induced by degraded nominal states in the conventional ESKF framework, this paper proposes an enhanced magnetic-inertial odometry method based on the invariant extended Kalman filtering (InEKF) framework. The attitude, velocity, and position are jointly formulated as elements of the Lie group of double direct spatial isometries ($SE_2(3)$), and a right-invariant error is defined exactly on this group. In addition, a scaled magnetic-field polynomial-model coefficient bias is introduced to compensate for errors in propagating the polynomial-model coefficients. The SINS error-state propagation and the magnetic-field model error-state propagation are rederived accordingly to construct the right-invariant error-state Kalman filter for magnetic-inertial odometry (RSE-KF-MIO).

The rest of this paper is organized as follows. Section II introduces the system and measurement models for MIO. Section III presents the conventional ESKF-based state estimation, which serves as the baseline method in this paper. Section IV describes the design of the RSE-KF-MIO based on the right-invariant error defined on the Lie group $SE_2(3)$. Section V reports the experimental setup and the results on public magnetometer-array datasets. Section VI concludes this paper.

## II. System and Measurement Models

Consider a rigid body moving indoors and equipped with an IMU and an array of $N$ magnetometer triads. The navigation frame is denoted by $n$, which is a local North—East—Down frame; the body frame is denoted by $b$ and is aligned with the IMU axes. The navigation states involve the rotation matrix $\mathbf{R}_b^n \in SO(3)$ that maps vectors from the $b$ frame to the $n$ frame, the velocity $\mathbf{v}^n \in \mathbb{R}^3$, and the position $\mathbf{p}^n \in \mathbb{R}^3$.

### A. Inertial Navigation Model

The SINS gyroscope and accelerometer measurements are modeled as follows:

$$\tilde{\boldsymbol{\omega}}^b = \boldsymbol{\omega}^b + \mathbf{b}_\omega + \mathbf{w}_\omega \tag{1}$$

$$\tilde{\mathbf{f}}^b = \left(\mathbf{R}_b^n\right)^\top \left(\mathbf{a}^n - \mathbf{g}^n\right) + \mathbf{b}_a + \mathbf{w}_a \tag{2}$$

where $\boldsymbol{\omega}^b$ denotes the true angular velocity expressed in the body frame, $\mathbf{a}^n$ denotes the true acceleration in the navigation frame. $\mathbf{g}^n = [0, 0, -9.82]^\top$ m/s$^2$ denotes the local gravity vector. $\mathbf{b}_\omega$ and $\mathbf{b}_a$ denote the gyroscope and accelerometer biases, respectively. $\mathbf{w}_\omega$ and $\mathbf{w}_a$ are the measurement noises of the gyroscope and accelerometer, modeled as zero-mean Gaussian white noise. The biases are modeled as random-walk processes:

$$\dot{\mathbf{b}}_a = \mathbf{w}_{b_a}, \qquad \dot{\mathbf{b}}_\omega = \mathbf{w}_{b_\omega} \tag{3}$$

where $\mathbf{w}_{b_a}$ and $\mathbf{w}_{b_\omega}$ are modeled as zero-mean Gaussian white noises.

For an SINS employing low-cost sensors and moving at moderate speeds, the effects of the transport rate, earth rotation and related terms can be neglected. The SINS mechanization equations are then expressed as follows:

$$\dot{\mathbf{R}}_b^n = \mathbf{R}_b^n \left[\boldsymbol{\omega}^b\right]_\times, \quad \dot{\mathbf{v}}^n = \mathbf{R}_b^n \mathbf{f}^b + \mathbf{g}^n, \quad \dot{\mathbf{p}}^n = \mathbf{v}^n \tag{4}$$

For a sampling period of $T_s = 1/f_s$, where $f_s$ is the SINS data update rate, the discrete-time recursion of the estimated SINS state is given by

$$\begin{aligned}\hat{\mathbf{x}}_{k+1}^{ins} &= \begin{bmatrix}\hat{\mathbf{p}}_{k+1}^n & \hat{\mathbf{v}}_{k+1}^n & \hat{\mathbf{R}}_{b_{k+1}}^n & \hat{\mathbf{b}}_{a,k+1} & \hat{\mathbf{b}}_{\omega,k+1}\end{bmatrix}^\top \\ &= f^{ins}\left(\hat{\mathbf{x}}_k^{ins}, \tilde{\boldsymbol{\omega}}_k^b, \tilde{\mathbf{f}}_k^b\right)\end{aligned} \tag{5}$$

where

$$f^{ins}\left(\hat{\mathbf{x}}_k^{ins}, \tilde{\boldsymbol{\omega}}_k^b, \tilde{\mathbf{f}}_k^b\right) = \begin{bmatrix} \hat{\mathbf{p}}_k + \hat{\mathbf{v}}_k T_s + \frac{1}{2}\hat{\mathbf{a}}_k^n T_s^2 \\ \hat{\mathbf{v}}_k + \hat{\mathbf{a}}_k^n T_s \\ \hat{\mathbf{R}}_{b_k}^n \exp\left(\left(\tilde{\boldsymbol{\omega}}_k^b - \hat{\mathbf{b}}_{\omega,k}\right)T_s\right) \\ \hat{\mathbf{b}}_{a,k} \\ \hat{\mathbf{b}}_{\omega,k} \end{bmatrix} \tag{6}$$

Here, $\hat{\mathbf{a}}_k^n = \hat{\mathbf{R}}_{b_k}^n(\tilde{\mathbf{f}}_k^b - \hat{\mathbf{b}}_{a,k}) + \mathbf{g}^n$ is the estimated acceleration in the $n$ frame, and $\exp(\cdot)$ denotes the exponential map on $\mathrm{SO}(3)$.

*B. Magnetic-Field Array Model*

At a spatial position $\mathbf{r} \in \mathbb{R}^3$, the magnetic field is modeled as $\mathbf{M}(\mathbf{r};\boldsymbol{\mu})$, where $\boldsymbol{\mu}$ denotes the vector of model parameters. If the magnetometer array moves in a static magnetic field with no free current, Maxwell's equations imply that, the magnetic field should be both curl free and divergence free [16]. The magnetic-field model therefore satisfies the following constraints:

$$\nabla_r \times \mathbf{M}(\mathbf{r};\boldsymbol{\mu}) = 0 \tag{7}$$

$$\nabla_r \cdot \mathbf{M}(\mathbf{r};\boldsymbol{\mu}) = 0 \tag{8}$$

Skog et al. [18] modeled the magnetic scalar potential as a polynomial of order ($l$+1) in the position components to satisfy the curl-free constraint given in (7); the magnetic field could be defined as the gradient of the scalar potential. The scalar potential function is expressed as follows:

$$\gamma(\mathbf{r};\boldsymbol{\mu}) = \mathbf{h}(\mathbf{r})^\top \boldsymbol{\mu} + c \tag{9}$$

where $\mathbf{h}(\mathbf{r})$ consists of terms $r_x^i r_y^j r_z^k$. $\forall i,j,k \in \mathbb{N}$ must satisfy $1 \le i+j+k \le m$, where $m = 1,2,\ldots,l+1$, and $c$ is a constant. Under these constraints, the number of combinations of $r_x^i r_y^j r_z^k$ is $L = (l+4)(l+3)(l+2)/6 - 1$. Therefore, the parameter vector is $\boldsymbol{\mu} \in \mathbb{R}^L$. Taking the gradient of the scalar potential $\gamma(\mathbf{r};\boldsymbol{\mu})$ yields an expression for the magnetic field:

$$\mathbf{M}(\mathbf{r};\boldsymbol{\mu}) = \nabla_r \gamma(\mathbf{r};\boldsymbol{\mu}) = \mathbf{A}(\mathbf{r})\boldsymbol{\mu} \tag{10}$$

where $\mathbf{A}(\mathbf{r}) = \nabla_r \mathbf{h}(\mathbf{r})^\top$.

Huang et al. [23] used the divergence-free constraint to eliminate redundant terms in $\boldsymbol{\mu}$, yielding the reparameterized polynomial model:

$$\mathbf{M}(\mathbf{r};\boldsymbol{\theta}) = \boldsymbol{\Phi}(\mathbf{r})\boldsymbol{\theta} \tag{11}$$

where $\boldsymbol{\Phi}(\mathbf{r}) \in \mathbb{R}^{3\times\kappa}$ is the polynomial-model regression matrix, $\boldsymbol{\theta} \in \mathbb{R}^\kappa$ is the polynomial-model coefficient. For an $l$-order polynomial model, the model has $\kappa = \dim(\theta) = l^2 + 4l + 3$ unknown parameters. Furthermore, the following expression for $\boldsymbol{\Phi}(\mathbf{r})$ when $l = 1$ was presented:

$$\boldsymbol{\Phi}(\mathbf{r}) = \begin{bmatrix} 0 & 0 & 1 & 0 & 0 & r_z & r_y & 2r_x \\ 0 & 1 & 0 & r_z & 2r_y & 0 & r_x & 0 \\ 1 & 0 & 0 & r_y & -2r_z & r_x & 0 & -2r_z \end{bmatrix} \tag{12}$$

Based on the magnetic-field polynomial-model, the following polynomial-model coefficient propagation equation in the body frame was derived in [23] :

$$\boldsymbol{\theta}_{k+1} = \mathbf{A}^\dagger \mathbf{B}(\boldsymbol{\psi}_k)\boldsymbol{\theta}_k + \mathbf{w}_k^{\boldsymbol{\theta}} \tag{13}$$

where $\mathbf{w}_k^{\boldsymbol{\theta}}$ is modeled as zero-mean Gaussian white noise with covariance $\mathbf{Q}_{\boldsymbol{\theta}} \in \mathbb{R}^{\kappa\times\kappa}$, $\mathbf{A}^\dagger$ denotes the Moore−Penrose inverse of $\mathbf{A}$. $\mathbf{A}$ is the regression matrix of the given locations expressed in the $b_{k+1}$ frame, and $\mathbf{B}$ is that of the same locations expressed in the $b_k$ frame and rotated into the $b_{k+1}$ frame. The expressions for $\mathbf{A}$ and $\mathbf{B}$ are as follows:

$$\mathbf{A} = \begin{bmatrix} \boldsymbol{\Phi}\left(\mathbf{r}_1^{b_{k+1}}\right) \\ \vdots \\ \boldsymbol{\Phi}\left(\mathbf{r}_S^{b_{k+1}}\right) \end{bmatrix}, \quad \mathbf{B}\left(\boldsymbol{\psi}_k\right) = \begin{bmatrix} \mathbf{R}_{b_k}^{b_{k+1}} \boldsymbol{\Phi}\left(\mathbf{r}_1^{b_k}\right) \\ \vdots \\ \mathbf{R}_{b_k}^{b_{k+1}} \boldsymbol{\Phi}\left(\mathbf{r}_S^{b_k}\right) \end{bmatrix} \tag{14}$$

where $\mathbf{R}_{b_k}^{b_{k+1}}$ denotes the rotation matrix that maps vectors from the $b_k$ frame to the $b_{k+1}$ frame. $\mathbf{r}_1,...,\mathbf{r}_S$ denote the position vector of given locations. The number of locations should be $S = \lceil \kappa / 3 \rceil$ to ensure that matrix $\mathbf{A}$ has full column rank. The position vectors of the location $s$ in the $b_k$ frame and $b_{k+1}$ frame are related by the following expression:

$$\mathbf{r}_s^{b_k} = \left(\mathbf{R}_{b_k}^{b_{k+1}}\right)^\top \mathbf{r}_s^{b_{k+1}} + \Delta\mathbf{p}_k^{b_k} \tag{15}$$

where $\Delta\mathbf{p}_k^{b_k} \in \mathbb{R}^3$ and $\mathbf{R}_{b_k}^{b_{k+1}}$ denotes the translation and rotation matrix from $b_k$ frame to the $b_{k+1}$ frame, which can be expressed as

$$\Delta\mathbf{p}_k^{b_k} = \left(\mathbf{R}_{b_k}^{n}\right)^\top \left(\mathbf{v}_k^n T_s + \mathbf{a}_k^n \frac{T_s^2}{2}\right) \tag{16}$$

$$\mathbf{R}_{b_k}^{b_{k+1}} = \left(\mathrm{Exp}\left(\Delta\phi_k\right)\right)^\top \tag{17}$$

where, $\Delta\phi_k = \boldsymbol{\omega}_k^{b_k} T_s$ is the orientation change from $b_k$ frame to the $b_{k+1}$ frame. A two-dimensional (2D) illustration of the geometric relationship between the two frames and the sensor platform used in [23] is shown in Fig. 1.

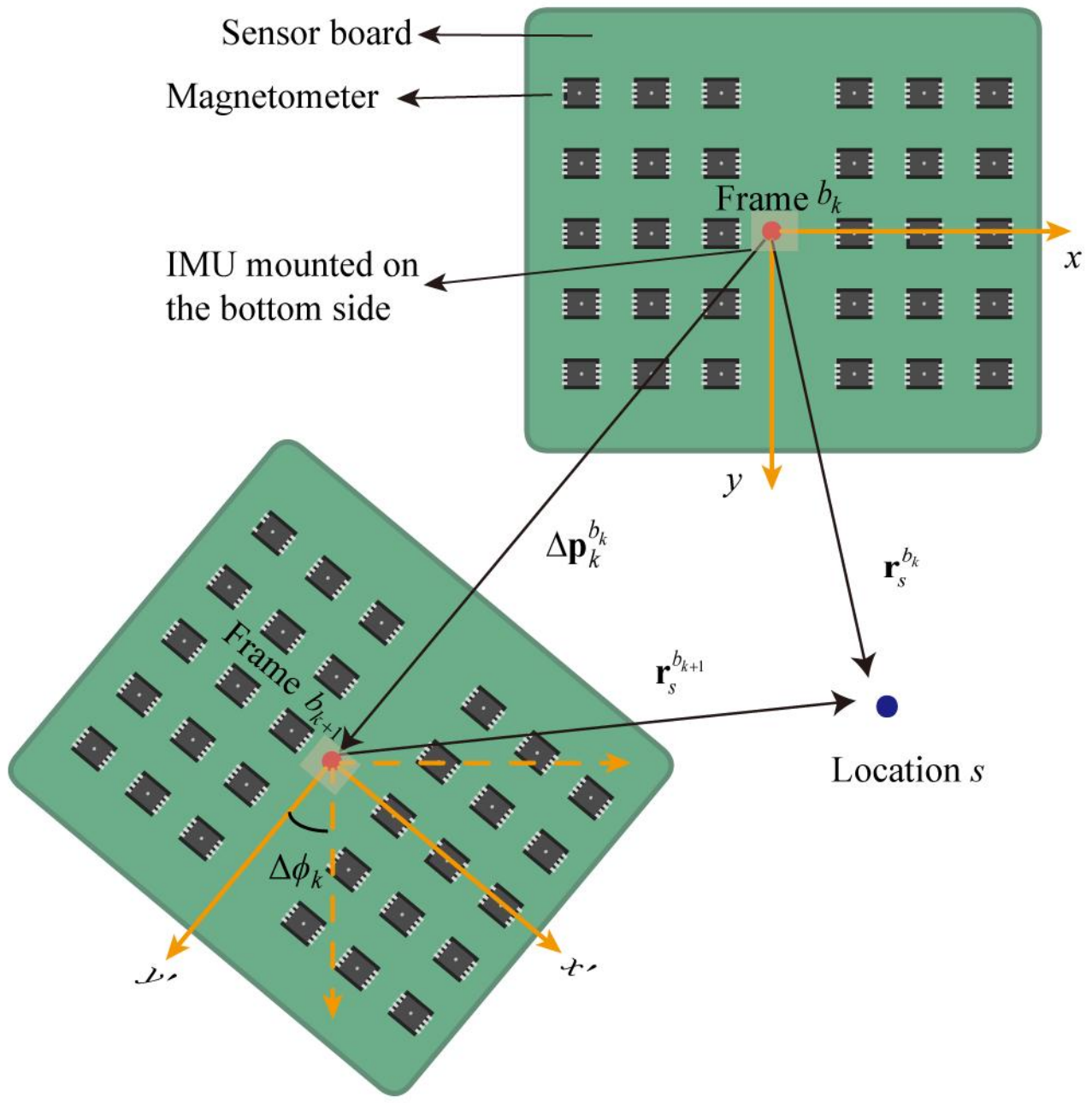

**Fig. 1.** 2-D illustration of the geometric relationship between the body frames at two consecutive times. The sensor board shown is a schematic depiction of the one used in [23], equipped with 30 PNI RM3100 magnetometers and an Osmium MIMU 4844 IMU mounted on the bottom side.

*C. Magnetometer-Array Measurement Model*

Based on the polynomial model (11), the measurement $\mathbf{z}_{m_i} \in \mathbb{R}^3$ from the $i$-th magnetometer located at $\mathbf{r}_{\mathrm{m}_i}$ in the body frame is given as follows:

$$\mathbf{z}_{m_i} = \mathbf{\Phi}(\mathbf{r}_{m_i})\mathbf{\theta} + \mathbf{e}_{m_i} \tag{18}$$

where $\mathbf{e}_{m_i}$ denotes measurement noise, modeled as zero-mean Gaussian white noise. With $N$ magnetic sensors in total, the measurement equation is given as follows:

$$\mathbf{z}_{\text{mag}} = \mathbf{\Phi}(\mathbf{r}_m)\mathbf{\theta} + \mathbf{e}_{\text{mag}} \tag{19}$$

where $\mathbf{\Phi}(\mathbf{r}_m) = \begin{bmatrix} \mathbf{\Phi}(\mathbf{r}_{m_i}) \\ \vdots \\ \mathbf{\Phi}(\mathbf{r}_{m_N}) \end{bmatrix} \in \mathbb{R}^{3N\times\kappa}$ .The covariance $\mathbf{R}_{\text{mag}}$ of the magnetic measurement noise $\mathbf{e}_{\text{mag}}$ is adaptively adjusted at $k$-th epoch :

$$\mathbf{R}_{\text{mag},k} = \hat{\sigma}_k^2 \mathbf{I}_{3N} \tag{20}$$

$$\hat{\sigma}_k^2 = \frac{1}{3N}\left\| \left( \mathbf{I}_{3N} - \mathbf{\Phi}(\mathbf{r}_{m,k})\mathbf{\Phi}^{\dagger}(\mathbf{r}_{m,k}) \right) \mathbf{z}_{\text{mag},k} \right\|^2 \tag{21}$$

*D. Position-Aiding Measurement Model*

Position aiding is optionally provided during an initial time window to constrain the IMU biases and stabilize the state estimates. This setting corresponds to scenarios in which GNSS or another positioning source is initially available. The measurement model is given as follows:

$$\mathbf{z}_{\text{pos}} = \mathbf{p} + \mathbf{e}_{\text{pos}}, \qquad \mathbf{e}_{\text{pos}} \sim \mathcal{N}(\mathbf{0}, \mathbf{R}_{\text{pos}}) \tag{22}$$

where $\mathbf{R}_{\text{pos}} = \sigma_p^2 \mathbf{I}_3$ . In the experiments, position aiding is enabled only during the first 60 s of each trajectory in the aided scenario and is disabled entirely in the unaided scenario.

## III. State Estimation using The ESKF

This section briefly introduces the ESKF framework employed by MAINS.

*A. Error-State Formulation*

The error-state vector in the conventional ESKF framework is defined as

$$\delta\mathbf{x} = \left[ \delta\mathbf{p}\ \delta\mathbf{v}\ \mathbf{\varepsilon}\ \delta\mathbf{b}_a\ \delta\mathbf{b}_\omega\ \delta\mathbf{\theta} \right]^\top \in \mathbb{R}^{15+\kappa} \tag{23}$$

where $\delta\mathbf{p} = \mathbf{p} - \hat{\mathbf{p}}$ , $\delta\mathbf{v} = \mathbf{v} - \hat{\mathbf{v}}$ , $\delta\mathbf{b}_a = \mathbf{b}_a - \hat{\mathbf{b}}_a$ , $\delta\mathbf{b}_\omega = \mathbf{b}_\omega - \hat{\mathbf{b}}_\omega$ and $\delta\mathbf{\theta} = \mathbf{\theta} - \hat{\mathbf{\theta}}$ are the additive errors in position, velocity, accelerometer bias, gyroscope bias and polynomial-model coefficient, respectively. $\mathbf{\varepsilon}$ is the small attitude error expressed in the $b$ frame, which satisfies the equation :

$$\mathbf{R}_b^n \simeq \hat{\mathbf{R}}_b^n \left( \mathbf{I}_3 + [\mathbf{\varepsilon}]_\times \right) \tag{24}$$

Based on measurement models (1), (2) and the SINS mechanization equations (4), the SINS error propagation equations are given by

$$\begin{aligned} \dot{\delta\mathbf{p}} &= \delta\mathbf{v} \\ \dot{\delta\mathbf{v}} &= -\hat{\mathbf{R}}_b^n \left[ \tilde{\mathbf{f}}^b - \hat{\mathbf{b}}_a \right]_\times \mathbf{\varepsilon} - \hat{\mathbf{R}}_b^n \delta\mathbf{b}_a + \hat{\mathbf{R}}_b^n \mathbf{w}_a \\ \dot{\mathbf{\varepsilon}} &= -\left[ \tilde{\mathbf{\omega}} - \hat{\mathbf{b}}_\omega \right]_\times \mathbf{\varepsilon} - \delta\mathbf{b}_\omega + \mathbf{w}_\omega \end{aligned} \tag{25}$$

Within the conventional ESKF framework, the SINS error propagation depends on the current nominal states and raw measurements: the velocity-error equation involves the estimated specific force $\tilde{\mathbf{f}}^b - \hat{\mathbf{b}}_a$ , whereas the attitude-error equation involves the estimated angular velocity $\tilde{\mathbf{\omega}} - \hat{\mathbf{b}}_\omega$ . Consequently, the linearized state-transition matrix $\mathbf{F}_k$ is time-varying and trajectory-dependent: it must be recomputed at every propagation step and is directly affected by the errors in the nominal specific-force and angular-rate estimates.

*B. Time and Measurement Updates*

The discrete-time error-state equation with a sampling period of $T_s$ in the conventional ESKF framework is expressed as follows:

$$\delta\mathbf{x}_{k+1} = \mathbf{F}_k^{\mathrm{E}}\delta\mathbf{x}_k + \mathbf{G}_k^{\mathrm{E}}\mathbf{w}_k \tag{26}$$

where $\delta\mathbf{x}_k = \left[\delta\mathbf{p}_k^\top\ \delta\mathbf{v}_k^\top\ \boldsymbol{\varepsilon}_k^\top\ \delta\mathbf{b}_{a,k}^\top\ \delta\mathbf{b}_{\omega,k}^\top\ \delta\boldsymbol{\theta}_k^\top\right]^\top$ and $\mathbf{w}_k = \left[\mathbf{w}_{a,k}^\top\ \mathbf{w}_{\omega,k}^\top\ \mathbf{w}_{b_a,k}^\top\ \mathbf{w}_{b_\omega,k}^\top\ \mathbf{w}_{\theta,k}^\top\right]^\top$ are the error state and the process noise at epoch $k$, respectively. $\mathbf{F}_k^{\mathrm{E}}$ is the error-state transition matrix, and $\mathbf{G}_k^{\mathrm{E}}$ is the noise-coupling matrix.

The covariance propagation is

$$\mathbf{P}_{k+1|k} = \mathbf{F}_k^{\mathrm{E}}\mathbf{P}_{k|k}\left(\mathbf{F}_k^{\mathrm{E}}\right)^\top + \mathbf{G}_k^{\mathrm{E}}\mathbf{Q}\left(\mathbf{G}_k^{\mathrm{E}}\right)^\top \tag{27}$$

where $\mathbf{Q}$ is the covariance of the process noise.

The measurement Jacobian $\mathbf{H}_{k+1}^{\mathrm{E}}$ follows directly from the measurement models (19) (22), and is expressed as

$$\mathbf{H}_{k+1}^{\mathrm{E}} = \begin{cases} \mathbf{H}_{\mathrm{mag},k+1}^{\mathrm{E}}, & \textit{Without Position Aiding} \\ \begin{bmatrix} \mathbf{H}_{\mathrm{mag},k+1}^{\mathrm{E}} \\ \mathbf{H}_{\mathrm{pos}}^{\mathrm{E}} \end{bmatrix}, & \textit{Position Aiding} \end{cases} \tag{28}$$

where $\mathbf{H}_{\mathrm{mag},k}^{\mathrm{E}} = \left[\mathbf{0}_{3N\times 15}\quad \boldsymbol{\Phi}(\mathbf{r}_{m,k})\right]$ is the magnetic-field measurement Jacobian, and $\mathbf{H}_{\mathrm{pos}}^{\mathrm{E}} = \left[\mathbf{I}_{3\times 3}\quad \mathbf{0}_{3\times(12+\kappa)}\right]$ is the position measurement Jacobian.

Given measurement residual $\delta\mathbf{z}_{k+1}$ and measurement covariance $\mathbf{R}_{k+1}$, the standard Kalman update is computed as

$$\begin{aligned} &\mathbf{K}_{k+1} \leftarrow \mathbf{P}_{k+1|k}\left(\mathbf{H}_{k+1}^{\mathrm{E}}\right)^\top\left(\mathbf{H}_{k+1}^{\mathrm{E}}\mathbf{P}_{k+1|k}\mathbf{H}_{k+1}^{\mathrm{E}\,\top} + \mathbf{R}_{k+1}\right)^{-1} \\ &\delta\hat{\mathbf{x}}_{k+1} \leftarrow \mathbf{K}_{k+1}\delta\mathbf{z}_{k+1} \\ &\mathbf{P}_{k+1|k+1} \leftarrow \left(\mathbf{I} - \mathbf{K}_{k+1}\mathbf{H}_{k+1}^{\mathrm{E}}\right)\mathbf{P}_{k+1|k} \end{aligned} \tag{29}$$

Then the update correction $\delta\hat{\mathbf{x}}_{k+1}$ is injected into the nominal state at epoch $k+1$ in a hybrid manner: additive corrections are applied to the position, velocity, biases, and polynomial-model coefficient, whereas a multiplicative correction is applied to the attitude matrix (or the quaternion).

## IV. Design of the RSE-KF-MIO

Fig. 2 illustrates the MIO scheme based on the RSE-KF-MIO. The IMU drives the nominal SINS state. The magnetometer array provides the magnetic measurements and constrains the polynomial-model coefficient state. Meanwhile, a coefficient-bias state, scaled by the scaling matrix, is introduced into the coefficient propagation to compensate for the model errors. These components constitute the right-invariant error-state-space model, which propagates the error state and updates it with the magnetic and optional position measurements. The estimated error state is then injected into the nominal state and reset. Note that short-term position measurements during the first 60 s are optionally used as an additional measurement, and both scenarios with and without position aiding are evaluated in this paper.

The remainder of this section is organized as follows. Section IV-A defines the right-invariant navigation error. Sections IV-B and IV-C derive the error-state propagation of the SINS states and of the magnetic-field model states, respectively, which are assembled into the complete right-invariant error-state-space model in Section IV-D. Finally, Section IV-E summarizes the RSE-KF-MIO algorithm.

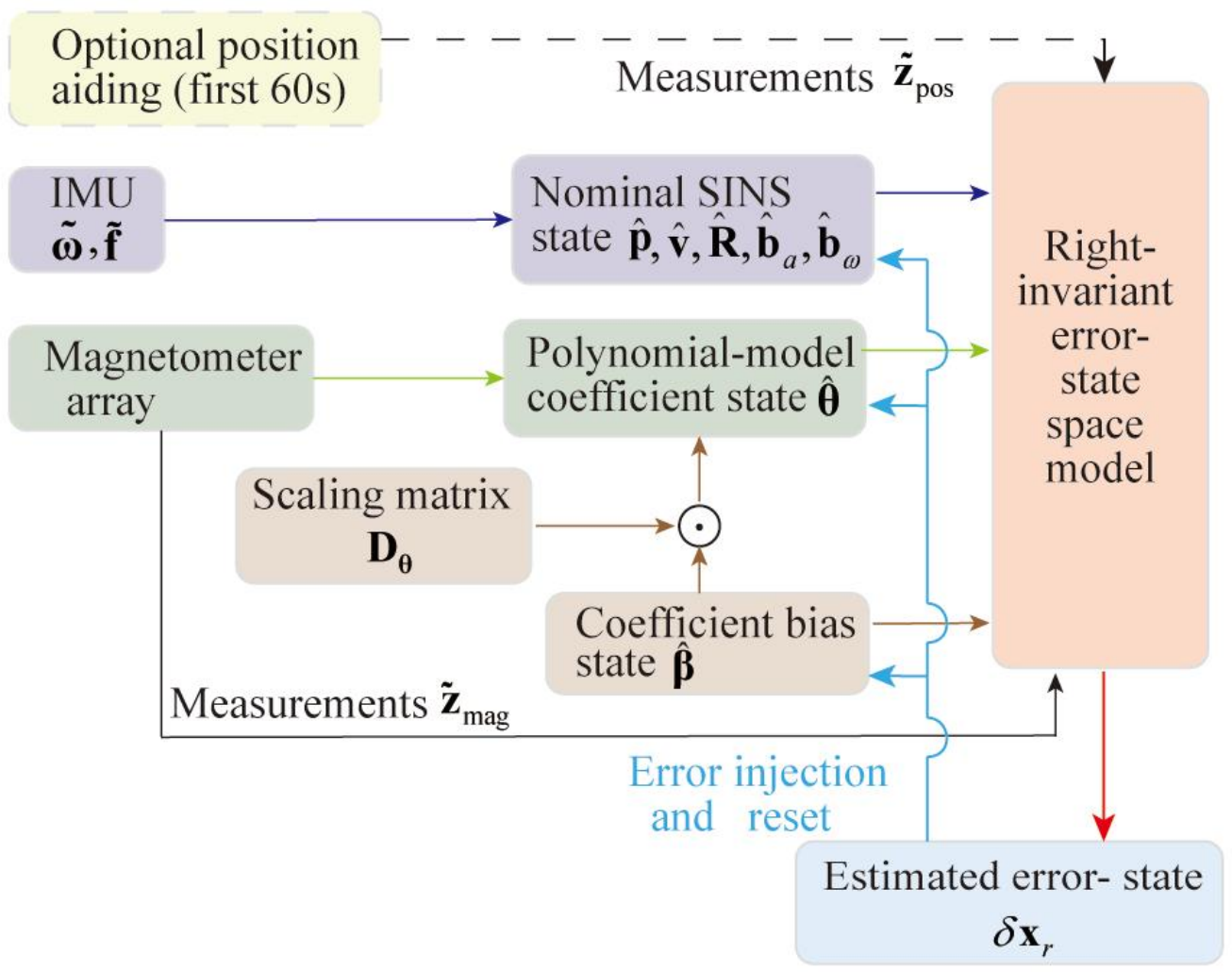

**Fig. 2.** Schematic of RSE-KF-MIO.

*A. Right-Invariant Error Definition*

Because the inertial navigation dynamics satisfy the group-affine condition [25], the attitude, velocity, and position are formulated as elements of Lie group SE2(3):

$$\mathbf{X} = \begin{bmatrix} \mathbf{R}_b^n & \mathbf{v}^n & \mathbf{p}^n \\ \mathbf{0}_{1\times3} & 1 & 0 \\ \mathbf{0}_{1\times3} & 0 & 1 \end{bmatrix} \in \mathrm{SE}_2(3) \tag{30}$$

The group operation is

$$\mathbf{X}_1\mathbf{X}_2 = \begin{bmatrix} \mathbf{R}_1\mathbf{R}_2 & \mathbf{R}_1\mathbf{v}_2 + \mathbf{v}_1 & \mathbf{R}_1\mathbf{p}_2 + \mathbf{p}_1 \\ \mathbf{0} & 1 & 0 \\ \mathbf{0} & 0 & 1 \end{bmatrix} \tag{31}$$

The right-invariant error between the estimate $\hat{\mathbf{X}}$ and the true state $\mathbf{X}$ is defined as

$$\boldsymbol{\eta}^r = \hat{\mathbf{X}}\mathbf{X}^{-1} = \begin{bmatrix} \hat{\mathbf{R}}_b^n \left(\mathbf{R}_b^n\right)^\top & \hat{\mathbf{v}} - \hat{\mathbf{R}}_b^n \left(\mathbf{R}_b^n\right)^\top \mathbf{v} & \hat{\mathbf{p}} - \hat{\mathbf{R}}_b^n \left(\mathbf{R}_b^n\right)^\top \mathbf{p} \\ \mathbf{0}_{1\times3} & 1 & 0 \\ \mathbf{0}_{1\times3} & 0 & 1 \end{bmatrix} \tag{32}$$

On the other hand, the matrix Lie group SE2(3) is related to its associated Lie algebra $\boldsymbol{\xi} = [\boldsymbol{\varphi}^\top \ \mathbf{v}_r^\top \ \mathbf{p}_r^\top]^\top \in \mathbb{R}^9$ through the exponential map:

$$\boldsymbol{\eta}^r = \exp\left(\boldsymbol{\xi}\right) = \begin{bmatrix} \exp(\boldsymbol{\varphi}) & \mathbf{J}_l(\boldsymbol{\varphi})\delta\mathbf{v}_r & \mathbf{J}_l(\boldsymbol{\varphi})\delta\mathbf{p}_r \\ \mathbf{0}_{1\times3} & 1 & 0 \\ \mathbf{0}_{1\times3} & 0 & 1 \end{bmatrix} \tag{33}$$

where $\exp(\boldsymbol{\varphi})$ is the SO(3) exponential map and $\mathbf{J}_l(\boldsymbol{\varphi})$ is the left Jacobian of SO(3), whose first-order approximations are given by

$$\exp(\boldsymbol{\varphi}) = \mathbf{I}_{3\times3} + [\boldsymbol{\varphi}]_\times \tag{34}$$

$$\mathbf{J}_l(\boldsymbol{\varphi}) = \mathbf{I}_{3\times3} + \frac{1}{2}[\boldsymbol{\varphi}]_\times \tag{35}$$

Equating the blocks of (32) and (33), and retaining only the first-order error terms yields the right-invariant errors of the attitude, velocity, and position expressed in the *n* frame:

$$\hat{\mathbf{R}}_b^n \left(\mathbf{R}_b^n\right)^\top = \mathbf{I}_3 + [\boldsymbol{\varphi}]_\times$$
$$\delta \mathbf{v}_r = \hat{\mathbf{v}} - \hat{\mathbf{R}}_b^n \left(\mathbf{R}_b^n\right)^\top \mathbf{v} \tag{36}$$
$$\delta \mathbf{p}_r = \hat{\mathbf{p}} - \hat{\mathbf{R}}_b^n \left(\mathbf{R}_b^n\right)^\top \mathbf{p}$$

The IMU biases and polynomial-model coefficient are not elements of Lie group SE2(3). They are retained as Euclidean additive errors $\delta\mathbf{b}_a = \mathbf{b}_a - \hat{\mathbf{b}}_a$ , $\delta\mathbf{b}_\omega = \mathbf{b}_\omega - \hat{\mathbf{b}}_\omega$ and $\delta\boldsymbol{\theta} = \boldsymbol{\theta} - \hat{\boldsymbol{\theta}}$ . In addition, this article introduces a polynomial-model coefficient bias term $\boldsymbol{\beta}$ , whose error is also defined as a Euclidean additive error $\delta\boldsymbol{\beta} = \boldsymbol{\beta} - \hat{\boldsymbol{\beta}}$ . Combining the invariant navigation error with these Euclidean additive errors, the complete error-state vector, which follows a hybrid invariant−Euclidean formulation, is given by

$$\delta \mathbf{x}_r = \left[\boldsymbol{\varphi}^\top \ \delta \mathbf{v}_r^\top \ \delta \mathbf{p}_r^\top \ \delta \mathbf{b}_a^\top \ \delta \mathbf{b}_\omega^\top \ \delta \boldsymbol{\theta}^\top \ \delta \boldsymbol{\beta}^\top\right]^\top \in \mathbb{R}^{15+2\kappa} \tag{37}$$

The corresponding nominal-state corrections are expressed as follows:

$$\hat{\mathbf{R}}_b^n \leftarrow \mathrm{Exp}\left(-[\boldsymbol{\varphi}]_\times\right)\hat{\mathbf{R}}_b^n$$
$$\hat{\mathbf{v}} \leftarrow \hat{\mathbf{v}} - \delta \mathbf{v}_r + \left[\hat{\mathbf{v}}\right]_\times \boldsymbol{\varphi} \tag{38}$$
$$\hat{\mathbf{p}} \leftarrow \hat{\mathbf{p}} - \delta \mathbf{p}_r + \left[\hat{\mathbf{p}}\right]_\times \boldsymbol{\varphi}$$

and

$$\hat{\mathbf{b}}_a \leftarrow \hat{\mathbf{b}}_a + \delta \mathbf{b}_a$$
$$\hat{\mathbf{b}}_\omega \leftarrow \hat{\mathbf{b}}_\omega + \delta \mathbf{b}_\omega$$
$$\hat{\boldsymbol{\theta}} \leftarrow \hat{\boldsymbol{\theta}} + \delta\boldsymbol{\theta} \tag{39}$$
$$\hat{\boldsymbol{\beta}} \leftarrow \hat{\boldsymbol{\beta}} + \delta\boldsymbol{\beta}$$

*B. SINS Error-State Propagation*

Differentiate the right-invariant error definitions in (36) with respect to time and substitute the SINS mechanization equations (4) and the IMU measurement models (1)–(2). Only the first-order terms in the error state are retained, giving the following continuous-time right-invariant error dynamics:

$$\dot{\boldsymbol{\varphi}} = \hat{\mathbf{R}}_b^n \left(\delta \mathbf{b}_\omega + \mathbf{w}_\omega\right) \tag{40}$$

$$\delta \dot{\mathbf{v}}_r = \left[\mathbf{g}^n\right]_\times \phi + \hat{\mathbf{R}}_b^n \left(\delta \mathbf{b}_a + \mathbf{w}_a\right) + \left[\hat{\mathbf{v}}\right]_\times \hat{\mathbf{R}}_b^n \left(\delta \mathbf{b}_\omega + \mathbf{w}_\omega\right) \tag{41}$$

$$\delta \dot{\mathbf{p}}_r = \delta \mathbf{v}_r + \left[\hat{\mathbf{p}}\right]_\times \hat{\mathbf{R}}_b^n \left(\delta \mathbf{b}_\omega + \mathbf{w}_\omega\right) \tag{42}$$

Combining the error dynamics (40)- (42) with the IMU bias dynamics (3) gives the continuous-time error-state equation of the SINS:

$$\delta \dot{\mathbf{x}}_r^{ins} = \mathbf{F}_r^{ins} \delta \mathbf{x}_r^{ins} + \mathbf{G}_r^{ins} \mathbf{w}^{ins} \tag{43}$$

where $\delta \mathbf{x}_r^{ins} = \left[\boldsymbol{\varphi}^\top \ \delta \mathbf{v}_r^\top \ \delta \mathbf{p}_r^\top \ \delta \mathbf{b}_a^\top \ \delta \mathbf{b}_\omega^\top\right]^\top \in \mathbb{R}^{15}$ is the error-state, $\mathbf{w}^{ins} = (\mathbf{w}_a^\top \ \mathbf{w}_\omega^\top \ \mathbf{w}_{b_a}^\top \ \mathbf{w}_{b_\omega}^\top)^\top$ is the process-noise vector, and the system matrices are given by

$$\mathbf{F}_r^{ins} = \begin{bmatrix} \mathbf{F}_0 & \mathbf{F}_1 \\ \mathbf{0}_{6\times 9} & \mathbf{0}_{6\times 6} \end{bmatrix}, \ \mathbf{G}_r^{ins} = \begin{bmatrix} \mathbf{G}_0 & \mathbf{0}_{9\times 6} \\ \mathbf{0}_{6\times 6} & \mathbf{I}_{6\times 6} \end{bmatrix} \tag{44}$$

where

$$\mathbf{F}_0 = \begin{bmatrix} \mathbf{0}_{3\times3} & \mathbf{0}_{3\times3} & \mathbf{0}_{3\times3} \\ \left[\mathbf{g}^n\right]_\times & \mathbf{0}_{3\times3} & \mathbf{0}_{3\times3} \\ \mathbf{0}_{3\times3} & \mathbf{I}_{3\times3} & \mathbf{0}_{3\times3} \end{bmatrix}, \quad \mathbf{F}_1 = \begin{bmatrix} \mathbf{0}_{3\times3} & \widehat{\mathbf{R}}_b^n \\ \widehat{\mathbf{R}}_b^n & \left[\hat{\mathbf{v}}\right]_\times \widehat{\mathbf{R}}_b^n \\ \mathbf{0}_{3\times3} & \left[\hat{\mathbf{p}}\right]_\times \widehat{\mathbf{R}}_b^n \end{bmatrix} \tag{45}$$

$$\mathbf{G}_0 = \begin{bmatrix} \mathbf{0}_{3\times3} & \widehat{\mathbf{R}}_b^n \\ \widehat{\mathbf{R}}_b^n & [\hat{\mathbf{v}}]_\times \widehat{\mathbf{R}}_b^n \\ \mathbf{0}_{3\times3} & [\hat{\mathbf{p}}]_\times \widehat{\mathbf{R}}_b^n \end{bmatrix} \tag{46}$$

Then, discretizing (43) with a sampling period of $T_s$ gives the discrete-time error-state equation of SINS:

$$\delta\mathbf{x}_{r,k+1}^{ins} = \mathbf{F}_k^{ins}\delta\mathbf{x}_{r,k}^{ins} + \mathbf{G}_k^{ins}\mathbf{w}_k^{ins} \tag{47}$$

where

$$\mathbf{F}_k^{ins} = \exp\left(\begin{bmatrix} \mathbf{F}_0 & \mathbf{F}_1 \\ \mathbf{0} & \mathbf{0} \end{bmatrix} T_s\right) = \begin{bmatrix} \mathbf{\Phi}_0 & \mathbf{\Psi}\mathbf{F}_1(\hat{\mathbf{x}}_k) \\ \mathbf{0} & \mathbf{I}_6 \end{bmatrix} \tag{48}$$

$$\mathbf{G}_k^{ins} = \begin{bmatrix} \mathbf{\Psi}\mathbf{G}_0\left(\hat{\mathbf{x}}_k\right) & \mathbf{0} \\ \mathbf{0} & \sqrt{T_s}\mathbf{I}_6 \end{bmatrix} \tag{49}$$

with $\mathbf{\Phi}_0$ and $\mathbf{\Psi}$ defined by

$$\mathbf{\Phi}_0 = \exp(\mathbf{F}_0 T_s) = \mathbf{I}_9 + \mathbf{F}_0 T_s + \frac{1}{2}\mathbf{F}_0^2 T_s^2 + \frac{1}{6}\mathbf{F}_0^3 T_s^3 + \cdots \tag{50}$$

$$\mathbf{\Psi} = \int_0^{T_s} e^{\mathbf{F}_0 \tau} d\tau \tag{51}$$

Since $\mathbf{F}_0$ is nilpotent, i.e., $\mathbf{F}_0^3 = \mathbf{0}$, both $\mathbf{\Phi}_0$ and $\mathbf{\Psi}$ have exact closed-form solutions:

$$\mathbf{\Phi}_0 = \mathbf{I}_{9\times9} + \mathbf{F}_0 T_s + \frac{1}{2}\mathbf{F}_0^2 T_s^2 = \begin{bmatrix} \mathbf{I}_3 & \mathbf{0}_{3\times3} & \mathbf{0}_{3\times3} \\ \left[\mathbf{g}^n\right]_\times T_s & \mathbf{I}_{3\times3} & \mathbf{0}_{3\times3} \\ \left[\mathbf{g}^n\right]_\times \frac{T_s^2}{2} & \mathbf{I}_3 T_s & \mathbf{I}_{3\times3} \end{bmatrix} \tag{52}$$

$$\mathbf{\Psi} = T_s \mathbf{I}_{9\times9} + \frac{1}{2}\mathbf{F}_0 T_s^2 + \frac{1}{6}\mathbf{F}_0^2 T_s^3 \tag{53}$$

Substituting (52) and (53) into (48) and (49)gives the final expressions of $\mathbf{F}_k^{ins}$ and $\mathbf{G}_k^{ins}$ :

$$\mathbf{F}_k^{ins} = \begin{bmatrix} \mathbf{I}_{3\times3} & \mathbf{0}_{3\times3} & \mathbf{0}_{3\times3} & \mathbf{0}_{3\times3} & \widehat{\mathbf{R}}_{b_k}^n T_s \\ \left[\mathbf{g}^n\right]_\times T_s & \mathbf{I}_{3\times3} & \mathbf{0}_{3\times3} & \widehat{\mathbf{R}}_{b_k}^n T_s & \left[\hat{\mathbf{v}}_k\right]_\times \widehat{\mathbf{R}}_{b_k}^n T_s \\ \left[\mathbf{g}^n\right]_\times \frac{T_s^2}{2} & T_s \mathbf{I}_{3\times3} & \mathbf{I}_{3\times3} & \widehat{\mathbf{R}}_{b_k}^n \frac{T_s^2}{2} & \begin{array}{c} [\hat{\mathbf{p}}_k]_\times \widehat{\mathbf{R}}_{b_k}^n T_s + \\ \left[\hat{\mathbf{v}}_k\right]_\times \widehat{\mathbf{R}}_{b_k}^n \frac{T_s^2}{2} \end{array} \\ \mathbf{0}_{3\times3} & \mathbf{0}_{3\times3} & \mathbf{0}_{3\times3} & \mathbf{I}_{3\times3} & \mathbf{0}_{3\times3} \\ \mathbf{0}_{3\times3} & \mathbf{0}_{3\times3} & \mathbf{0}_{3\times3} & \mathbf{0}_{3\times3} & \mathbf{I}_{3\times3} \end{bmatrix} \tag{54}$$

$$
\mathbf{G}_k^{ins} = \begin{bmatrix} \mathbf{0}_{3\times3} & \widehat{\mathbf{R}}_{b_k}^{n} T_s & \mathbf{0}_{3\times3} & \mathbf{0}_{3\times3} \\ \widehat{\mathbf{R}}_{b_k}^{n} T_s & \left[\hat{\mathbf{v}}_k\right]_\times \widehat{\mathbf{R}}_{b_k}^{n} T_s & \mathbf{0}_{3\times3} & \mathbf{0}_{3\times3} \\ \widehat{\mathbf{R}}_{b_k}^{n} \dfrac{T_s^2}{2} & \begin{array}{c} \left[\hat{\mathbf{v}}_k\right]_\times \widehat{\mathbf{R}}_{b_k}^{n} \dfrac{T_s^2}{2} \\ +[\hat{\mathbf{p}}_k]_\times \widehat{\mathbf{R}}_{b_k}^{n} T_s \end{array} & \mathbf{0}_{3\times3} & \mathbf{0}_{3\times3} \\ \mathbf{0}_{3\times3} & \mathbf{0}_{3\times3} & \sqrt{T_s}\mathbf{I}_{3\times3} & \mathbf{0}_{3\times3} \\ \mathbf{0}_{3\times3} & \mathbf{0}_{3\times3} & \mathbf{0}_{3\times3} & \sqrt{T_s}\mathbf{I}_3 \end{bmatrix} \tag{55}
$$

where the expansions are truncated according to the integration depth of each error state: the velocity-error blocks retain only the terms linear in $T_s$ , the position-error blocks retain terms up to quadratic in $T_s$ , and all higher-order terms are discarded.

The right-invariant formulation offers two advantages over the conventional ESKF:

1) *Trajectory-independent propagation of the drift core*. The propagation of the navigation-error triplet $(\boldsymbol{\varphi}, \delta\mathbf{v}_r, \delta\mathbf{p}_r)$ is governed by $\boldsymbol{\Phi}_0$ , which is constant and involves neither state estimates nor inertial measurements. The bias-coupling terms, though dependent on the nominal state, are non-dominant and their influence remains limited.

2) *Closed-form state transition*. Owing to the nilpotency of $\mathbf{F}_0$ i.e., $\mathbf{F}_0^3 = \mathbf{0}$ , $\mathbf{F}_k^{ins}$ admits an exact closed-form expression, whereas the ESKF transition matrix depends on the trajectory and is commonly limited to a first-order approximation.

*C. Magnetic-Field Polynomial-Model Error-State Propagation*

The propagation for the local magnetic-field polynomial-model coefficient is given in (13). It relies on the assumption that the applicable region of the local magnetic-field model covers the spatial neighborhoods of the platform positions at consecutive epochs. This assumption fails when the inter-epoch displacement of the platform is large, as in rapid motion, or when the magnetic anomalies vary strongly over space, in which case the field at the current position is no longer adequately represented by the previously identified local model. Continuing to propagate the polynomial-model coefficient with the original equation then accumulates model errors. To absorb this mismatch, a time-varying bias $\boldsymbol{\beta}$ is introduced into the polynomial-model coefficient propagation equation, and the revised propagation equation is expressed as follows:

$$
\boldsymbol{\theta}_{k+1} = \mathbf{A}^{\dagger}\mathbf{B}(\boldsymbol{\psi}_k)\boldsymbol{\theta}_k + \mathbf{D}_{\theta}\boldsymbol{\beta}_k + \mathbf{w}_k^{\theta} \tag{56}
$$

where $\mathbf{D}_{\boldsymbol{\theta}}\boldsymbol{\beta}_k$ is the newly introduced polynomial-model coefficient bias term. $\boldsymbol{\beta}_k$ is the polynomial-model coefficient bias state, and $\mathbf{D}_{\boldsymbol{\theta}}$ is the coefficient scaling matrix, which maps the dimensionless bias state $\boldsymbol{\beta}_k$ to the numerical scales of the different polynomial-model coefficients. The coefficient bias state follows a random-walk model:

$$
\boldsymbol{\beta}_{k+1} = \boldsymbol{\beta}_k + \sqrt{T_s}\mathbf{w}_k^{\beta}, \quad \mathbf{w}_k^{\beta} \sim \mathcal{N}(\mathbf{0}, \mathbf{Q}_{\beta}) \tag{57}
$$

where $\mathbf{Q}_{\beta} = \sigma_{\beta}^2 \mathbf{I}_{\kappa}$ .

To balance the compensation strength across the model-coefficient channels, the scaling matrix $\mathbf{D}_{\boldsymbol{\theta}}$ is determined by covariance matching. Let $\mathbf{Q}_{\Delta\boldsymbol{\theta}}$ denote the covariance matrix of the term $\mathbf{D}_{\boldsymbol{\theta}}\boldsymbol{\beta}_k$ , then

$$
\mathbf{Q}_{\Delta\boldsymbol{\theta}} = \mathbf{D}_{\boldsymbol{\theta}}\mathbf{Q}_{\boldsymbol{\beta}}\mathbf{D}_{\boldsymbol{\theta}}^{\top} = \sigma_{\beta}^2 \mathbf{D}_{\boldsymbol{\theta}}\mathbf{D}_{\boldsymbol{\theta}}^{\top} \tag{58}
$$

The scaling matrix is determined by requiring the compensation term to inherit the statistical scale of the original coefficient process noise, i.e.,

$$
\mathbf{Q}_{\Delta\boldsymbol{\theta}} \propto \mathbf{Q}_{\boldsymbol{\theta}} \tag{59}
$$

The simplest choice satisfying this constraint is

$$
\mathbf{D}_{\boldsymbol{\theta}}\mathbf{D}_{\boldsymbol{\theta}}^{\top} = \mathbf{Q}_{\boldsymbol{\theta}} \tag{60}
$$

so that

$$\mathbf{D}_{\boldsymbol{\theta}} = \operatorname{diag}\left(\sigma_{\theta,1},\ldots,\sigma_{\theta,\kappa}\right) \tag{61}$$

where $\sigma_{\theta,i}$ is the standard deviation of the $i$-th coefficient channel. Each channel is thus compensated in proportion to its own noise level, while the single parameter $\sigma_{\beta}$ governs the overall compensation strength.

To obtain the error dynamics of the polynomial-model coefficient, the propagation equation (56) is linearized about the nominal coefficient state $\hat{\boldsymbol{\theta}}_k$ and the nominal inter-epoch relative pose $\left[\Delta\hat{\mathbf{p}}_k^{b\top} \quad \Delta\hat{\phi}_k^{\top}\right]^{\top}$, with only the first-order terms retained:

$$\delta\boldsymbol{\theta}_{k+1} = \mathbf{A}^{\dagger}\mathbf{B}\delta\boldsymbol{\theta}_k + \mathbf{A}^{\dagger}\mathbf{J}_1\delta\Delta\mathbf{p}_k^{b_k} + \mathbf{A}^{\dagger}\mathbf{J}_2\delta\Delta\phi_k + \mathbf{D}_{\boldsymbol{\theta}}\delta\boldsymbol{\beta}_k + \mathbf{w}_k^{\theta} \tag{62}$$

where $\delta\Delta\mathbf{p}_k^{b_k}$ and $\delta\Delta\phi_k$ are the errors of the translation and orientation change from the $b_k$ frame to the $b_{k+1}$ frame, and

$$\mathbf{J}_1 = \frac{d}{d\Delta\mathbf{p}_k^{b}}\left(\mathbf{B}\left(\hat{\boldsymbol{\psi}}_k\right)\boldsymbol{\theta}_k\right) \tag{63}$$

$$\mathbf{J}_2 = \frac{d}{d\Delta\phi_k}\left(\mathbf{B}\left(\hat{\boldsymbol{\psi}}_k\right)\boldsymbol{\theta}_k\right) \tag{64}$$

Here, $\delta\Delta\mathbf{p}_k^{b_k}$ and $\delta\Delta\phi_k$ are re-expressed in terms of the right-invariant errors. For $\delta\Delta\mathbf{p}_k^{b}$, using the corresponding translation expression (16) gives

$$\begin{aligned}
\delta\Delta\mathbf{p}_k^{b_k} &= \Delta\mathbf{p}_k^{b_k} - \Delta\hat{\mathbf{p}}_k^{b_k} \\
&= \left(\mathbf{R}_{b_k}^n\right)^{\top}\left(\mathbf{v}_k^n T_s + \mathbf{a}_k^n\frac{T_s^2}{2}\right) - \left(\hat{\mathbf{R}}_{b_k}^n\right)^{\top}\left(\hat{\mathbf{v}}_k^n T_s + \hat{\mathbf{a}}_k^n\frac{T_s^2}{2}\right) \\
&\simeq -\frac{T_s^2}{2}\left(\hat{\mathbf{R}}_{b_k}^n\right)^{\top}\left[\mathbf{g}^n\right]_{\times}\boldsymbol{\varphi}_k - T_s\hat{\mathbf{R}}_{b_k}^{n\top}\delta\mathbf{v}_{r,k} - \frac{T_s^2}{2}\left(\delta\mathbf{b}_{a,k} + \mathbf{w}_{a,k}\right)
\end{aligned} \tag{65}$$

Similarly, $\delta\Delta\phi_k$ is expressed as

$$\delta\Delta\phi_k = \boldsymbol{\omega}_k^{b_k}T_s - \hat{\boldsymbol{\omega}}_k^{b_k}T_s = -T_s\delta\mathbf{b}_{\omega,k} - T_s\mathbf{w}_{\omega,k} \tag{66}$$

Combining (62)-(66) yields

$$\delta\boldsymbol{\theta}_{k+1} = \mathbf{F}_k^{\theta}\begin{bmatrix}\delta\mathbf{x}_{r,k}^{ins} \\ \delta\boldsymbol{\theta}_k \\ \delta\boldsymbol{\beta}_k\end{bmatrix} + \mathbf{G}_k^{\theta}\begin{bmatrix}\mathbf{w}_k^{ins} \\ \mathbf{w}_k^{\theta} \\ \mathbf{w}_k^{\beta}\end{bmatrix} \tag{67}$$

where

$$\mathbf{F}_k^{\theta} = \left[\mathbf{A}^{\dagger}\left[\mathbf{B}(\hat{\psi}_k) \;\; \mathbf{J}_1 \;\; \mathbf{J}_2\right]\mathbf{C}\left(\hat{\mathbf{x}}_k\right) \;\; \mathbf{D}_{\boldsymbol{\theta}}\right] \tag{68}$$

$$\mathbf{G}_k^{\theta} = \left[-\frac{T_s^2}{2}\mathbf{A}^{\dagger}\mathbf{J}_1 \quad -\mathbf{A}^{\dagger}\mathbf{J}_2 T_s \quad \mathbf{0}_{\kappa\times 3} \quad \mathbf{0}_{\kappa\times 3} \quad \mathbf{I}_{\kappa\times\kappa} \quad \mathbf{0}_{\kappa\times\kappa}\right] \tag{69}$$

and

$$\mathbf{C}\left(\mathbf{x}_k\right) = \begin{bmatrix}
\mathbf{0}_{\kappa\times 3} & \mathbf{0}_{\kappa\times 3} & \mathbf{0}_{\kappa\times 3} & \mathbf{0}_{\kappa\times 3} & \mathbf{0}_{\kappa\times 3} & \mathbf{I}_{\kappa\times\kappa} \\
\frac{-T_s^2}{2}\left(\hat{\mathbf{R}}_{b_k}^n\right)^{\top}\left[\mathbf{g}^n\right]_{\times} & -T_s\left(\hat{\mathbf{R}}_{b_k}^n\right)^{\top} & \mathbf{0}_{3\times 3} & \frac{-T_s^2}{2}\mathbf{I}_{3\times 3} & \mathbf{0}_{3\times 3} & \mathbf{0}_{3\times\kappa} \\
\mathbf{0}_{3\times 3} & \mathbf{0}_{3\times 3} & \mathbf{0}_{3\times 3} & \mathbf{0}_{3\times 3} & -T_s\mathbf{I}_{3\times 3} & \mathbf{0}_{3\times\kappa}
\end{bmatrix} \tag{70}$$

Based on (57), the error-state propagation equation for the polynomial-model coefficient bias is

$$\delta\boldsymbol{\beta}_{k+1} = \delta\boldsymbol{\beta}_k + \sqrt{T_s}\,\mathbf{w}_k^{\boldsymbol{\beta}} \tag{71}$$

*D. Complete Right-Invariant Error-State-Space Model*

Combining the SINS error-state propagation (47) with the magnetic-field polynomial-model error-state propagation (62)and (71) gives the complete right-invariant error-state propagation equation:

$$\delta\mathbf{x}_{r,k+1} = \mathbf{F}_k \delta\mathbf{x}_{r,k} + \mathbf{G}_k \mathbf{w}_k \quad (72)$$

where

$$\delta\mathbf{x}_{r,k} = \begin{bmatrix} \delta\mathbf{x}_{r,k}^{ins} \\ \delta\boldsymbol{\theta}_k \\ \delta\boldsymbol{\beta}_k \end{bmatrix}, \quad \mathbf{F}_k = \begin{bmatrix} \mathbf{F}_k^{ins} & \mathbf{0}_{15\times\kappa} & \mathbf{0}_{15\times\kappa} \\ & \mathbf{F}_k^{\theta} & \\ \mathbf{0}_{\kappa\times15} & \mathbf{0}_{\kappa\times\kappa} & \mathbf{I}_{\kappa\times\kappa} \end{bmatrix} \quad (73)$$

$$\mathbf{G}_k = \begin{bmatrix} \mathbf{G}_k^{ins} & \mathbf{0}_{15\times\kappa} & \mathbf{0}_{15\times\kappa} \\ & \mathbf{G}_k^{\theta} & \\ \mathbf{0}_{\kappa\times12} & \mathbf{0}_{\kappa\times\kappa} & \sqrt{T_s}\mathbf{I}_{\kappa\times\kappa} \end{bmatrix}, \quad \mathbf{w}_k = \begin{bmatrix} \mathbf{w}_k^{ins} \\ \mathbf{w}_k^{\theta} \\ \mathbf{w}_k^{\beta} \end{bmatrix} \quad (74)$$

Here, $\mathbf{w}_k$ is a zero-mean Gaussian white sequence with covariance:

$$\mathbf{Q}_k = \mathbf{Cov}(\mathbf{w}_k) = \text{blkdiag}\left(\boldsymbol{\Sigma}_a \ \boldsymbol{\Sigma}_g \ \boldsymbol{\Sigma}_{b_a} \ \boldsymbol{\Sigma}_{b_w} \ \boldsymbol{\Sigma}_\theta \ \boldsymbol{\Sigma}_\beta\right) \quad (75)$$

where $\text{blkdiag}(\bullet)$ denotes the operator that constructs a block-diagonal matrix, and $\boldsymbol{\Sigma}_{(\cdot)}$ denotes the covariance matrix of the corresponding noise component.

For the measurement update, the magnetometer measurements in error form give

$$\delta\mathbf{z}_{\text{mag},k+1} = \mathbf{z}_{\text{mag},k+1} - \boldsymbol{\Phi}(\mathbf{r}_{m,k+1})\hat{\boldsymbol{\theta}}_{k+1} = \boldsymbol{\Phi}(\mathbf{r}_{m,k+1})\delta\boldsymbol{\theta}_{k+1} + \mathbf{e}_{\text{mag},k+1} \quad (76)$$

When position-aiding measurements are available, the position error measurement is

$$\delta\mathbf{z}_{\text{pos},k+1} = \mathbf{z}_{\text{pos},k+1} - \hat{\mathbf{p}}_{k+1} = \mathbf{p}_{k+1} - \hat{\mathbf{p}}_{k+1} + \mathbf{e}_{\text{pos},k+1} \quad (77)$$

To express (77) in terms of the right-invariant errors, right-invariant position error definition in (36) is rearranged as

$$\mathbf{p} = \mathbf{R}_b^n\left(\hat{\mathbf{R}}_b^n\right)^{\top}(\hat{\mathbf{p}} - \delta\mathbf{p}_r) \simeq (\mathbf{I}_3 - [\boldsymbol{\varphi}]_\times)(\hat{\mathbf{p}} - \delta\mathbf{p}_r) \simeq \hat{\mathbf{p}} - \delta\mathbf{p}_r + \left[\hat{\mathbf{p}}\right]_\times \boldsymbol{\varphi} \quad (78)$$

Substituting (78) into (77) gives

$$\delta\mathbf{z}_{\text{pos},k+1} = -\delta\mathbf{p}_{r,k+1} + \left[\hat{\mathbf{p}}_{k+1}\right]_\times \boldsymbol{\varphi}_{k+1} + \mathbf{e}_{\text{pos},k+1} \quad (79)$$

Combining the magnetic measurement (76) and optional position measurement (79) gives the measurement-update equation:

$$\delta\mathbf{z}_{k+1} = \mathbf{H}_{k+1}\delta\mathbf{x}_{r,k+1} + \mathbf{e}_{k+1} \quad (80)$$

$$\delta\mathbf{z}_{k+1} = \begin{cases} \delta\mathbf{z}_{\text{mag},k+1}, & \text{Without Position Aiding} \\ \begin{bmatrix} \delta\mathbf{z}_{\text{mag},k+1} \\ \delta\mathbf{z}_{\text{pos},k+1} \end{bmatrix}, & \text{With Position Aiding} \end{cases} \quad (81)$$

$$\mathbf{H}_{k+1} = \begin{cases} \mathbf{H}_{\text{mag},k+1}, & \text{Without Position Aiding} \\ \begin{bmatrix} \mathbf{H}_{\text{mag},k+1} \\ \mathbf{H}_{\text{pos},k+1} \end{bmatrix}, & \text{With Position Aiding} \end{cases} \quad (82)$$

$$\mathbf{R}_{k+1} = \mathbf{Cov}(\mathbf{e}_{k+1}) = \begin{cases} \mathbf{R}_{\text{mag},k+1}, & \text{Without Position Aiding} \\ \text{blkdiag}\left(\mathbf{R}_{\text{mag},k+1}, \mathbf{R}_{\text{pos},k+1}\right), & \text{With Position Aiding} \end{cases} \quad (83)$$

and

$$\mathbf{H}_{\text{mag},k+1} = \left[\mathbf{0}_{3N\times15} \quad \boldsymbol{\Phi}(\mathbf{r}_{m,k+1}) \quad \mathbf{0}_{3N\times\kappa}\right] \quad (84)$$

$$\mathbf{H}_{\text{pos},k+1} = \left[\left[\hat{\mathbf{p}}_{k+1}\right]_\times \quad \mathbf{0}_{3\times3} \quad -\mathbf{I}_{3\times3} \quad \mathbf{0}_{3\times(6+2\kappa)}\right] \quad (85)$$

### E. Summary of the RSE-KF-MIO Algorithm

The complete description of the proposed RSE-KF-MIO is summarized in Algorithm 1.

---

**Algorithm 1:** RSE-KF-MIO

---

**Input:** IMU sequence $\left\{\tilde{\mathbf{f}}_k^b, \tilde{\boldsymbol{\omega}}_k^b\right\}_{k=0}^{N}$, step size $T_s$, magnetometer-array measurements $\left\{\tilde{\mathbf{z}}_{\text{mag}}\right\}_{k=0}^{N}$, position-aiding measurements $\left\{\tilde{\mathbf{z}}_{\text{pos}}\right\}_{k=0}^{60/T_s-1}$.

**Output:** $\left\{\hat{\mathbf{x}}_{k|k}, \mathbf{P}_{k|k}\right\}_{k=1}^{N}$

**Initialization**: nominal-state estimate $\hat{\mathbf{x}}_{0|0}$, covariance matrix $\mathbf{P}_{0|0}$

**For** $k=0$ to $k=N-1$ do

**Nominal SINS state propagation**

$$\hat{\mathbf{x}}_{k+1|k}^{ins} \leftarrow f\left(\hat{\mathbf{x}}_k^{ins}, \tilde{\mathbf{f}}_k^b, \tilde{\boldsymbol{\omega}}_k^b\right) \text{ [cf. (6)]}$$

**Nominal polynomial-model state propagation**

$$\hat{\boldsymbol{\theta}}_{k+1|k} \leftarrow \mathbf{A}^{\dagger}\mathbf{B}(\hat{\boldsymbol{\psi}}_k)\hat{\boldsymbol{\theta}}_k + \mathbf{D}_{\theta}\hat{\boldsymbol{\beta}}_k$$

$$\hat{\boldsymbol{\beta}}_{k+1|k} \leftarrow \hat{\boldsymbol{\beta}}_k$$

**Error-state covariance propagation**

$$\mathbf{P}_{k+1|k} = \mathbf{F}_k\mathbf{P}_{k|k}\mathbf{F}_k^{\top} + \mathbf{G}_k\mathbf{Q}_k\mathbf{G}_k^{\top}$$

**Measurement update**:

$$\delta\mathbf{z}_{k+1} = \begin{cases} \tilde{\mathbf{z}}_{\text{mag},k+1} - \boldsymbol{\Phi}\hat{\boldsymbol{\theta}}_{k+1|k}, & \text{Without Position Aiding} \\ \begin{bmatrix} \tilde{\mathbf{z}}_{\text{mag},k+1} - \boldsymbol{\Phi}\hat{\boldsymbol{\theta}}_{k+1|k} \\ \tilde{\mathbf{z}}_{\text{pos},k+1} - \hat{\mathbf{p}}_{k+1|k} \end{bmatrix}, & \text{With Position Aiding} \end{cases}$$

$$\mathbf{S}_{k+1} \leftarrow \mathbf{H}_{k+1}\mathbf{P}_{k+1|k}\mathbf{H}_{k+1}^{\top} + \mathbf{R}_{k+1}$$

$$\mathbf{K}_{k+1} \leftarrow \mathbf{P}_{k+1|k}\mathbf{H}_{k+1}^{\top}\mathbf{S}_{k+1}^{-1}$$

$$\delta\mathbf{x}_{r,k+1} \leftarrow \mathbf{K}_{k+1}\delta\mathbf{z}_{k+1}$$

$$\mathbf{P}_{k+1|k+1} = \left(\mathbf{I} - \mathbf{K}_{k+1}\mathbf{H}_{k+1}\right)\mathbf{P}_{k+1|k}\left(\mathbf{I} - \mathbf{K}_{k+1}\mathbf{H}_{k+1}\right)^{\top} + \mathbf{K}_{k+1}\mathbf{R}_{k+1}\mathbf{K}_{k+1}^{\top}$$

**Error correction**:

$$\hat{\mathbf{x}}_{k+1|k+1} \leftarrow \hat{\mathbf{x}}_{k+1|k} \oplus \delta\mathbf{x}_{r,k+1}, \text{ [cf. (38) and (39)]}$$

**Error reset**:

$$\delta\mathbf{x}_{r,k+1} \leftarrow \mathbf{0}$$

**End for**

---

## V. Experimental Setup and Results

### A. Datasets and Experimental Parameter Settings

The experiments are conducted on the public magnetometer-array dataset released in [23]. The data were collected using a sensor array comprising 30 magnetometer triads and an IMU. The magnetometer array and IMU were mounted on a planar board measuring 235 mm × 345 mm. A schematic of the sensor platform is shown in Fig. 1. The ground-truth trajectory of the platform was measured using a camera-based motion-tracking system. Eight motion experiments were conducted at different board heights and inclination angles relative to the ground. Table I summarizes the configurations of the eight experiments.

A first-order polynomial magnetic-field model with $\kappa = 8$ coefficients is used. The square and rectangular sensor configurations are used, comprising 5 and 30 magnetometers, respectively. MAINS and the proposed RSE-KF-MIO share the basic filter parameters. All error states are initialized as zero vectors. The nominal position and attitude are initialized from the first ground-truth frame; velocity and IMU biases are initialized to zero; and the magnetic-field coefficients are estimated from the first magnetic measurement by least squares. The initial error-state standard deviations are $\sigma_{\boldsymbol{\varepsilon}} = 10^{\circ}$, $\sigma_{\delta\mathbf{v}} = 1\ \text{m/s}$, $\sigma_{\delta\mathbf{p}} = 0.1\ \text{m}$, $\sigma_{\delta\mathbf{b}_a} = 0.5\ \text{m/s}^2$, $\sigma_{\delta\mathbf{b}_\omega} = 0.5^{\circ}/\text{s}$, $\sigma_{\delta\boldsymbol{\theta}} = 0.2$, $\sigma_{\delta\boldsymbol{\beta}} = 0.1$. The process-noise standard deviations are $\sigma_{\mathbf{w}_a} = 0.05\ \text{m/s}^2$, $\sigma_{\mathbf{w}_\omega} = 0.5^{\circ}/\text{s}$, $\sigma_{\mathbf{w}_{b_a}} = 10^{-7}\,\text{m/s}^{5/2}$, $\sigma_{\mathbf{w}_{b_\omega}} = 10^{-7}\ {}^{\circ}/\text{s}^{3/2}$, $\sigma_{\mathbf{w}_\theta} = \left[0.0005\ 0.0003\ 0.001\ 0.01\ 0.01\ 0.02\ 0.02\ 0.01\right]^{\top}$, $\sigma_{\mathbf{w}_\beta} = 0.001$. The scaling matrix is $\mathbf{D}_{\boldsymbol{\theta}} = \text{diag}\left(\sigma_{\mathbf{w}_\theta}\right)$. The position measurement-noise covariance is $\mathbf{R}_{\text{pos}} = (0.01)^2\mathbf{I}_3\ \text{m}^2$. The magnetic-array measurement-noise covariance is $\mathbf{R}_{\text{mag},k} = \hat{\sigma}^2_{\text{mag},k}\mathbf{I}_{3N}$, where the standard deviation $\hat{\sigma}_{\text{mag}}$ is adaptively estimated using (21) at each epoch.

TABLE I
DATASETS INFORMATION

| Data sequence [1] | LP-1 | LP-2 | LP-3 | NP-1 | NP-2 | NP-3 | NT-1 | NT-2 |
|---|---|---|---|---|---|---|---|---|
| Trajectory length (m) | 138.72 | 167.07 | 194.41 | 136.23 | 134.66 | 137.76 | 164.62 | 137.87 |
| Trajectory duration (s) | 272 | 286 | 332 | 177 | 165 | 154 | 185 | 151 |
| Average height (m) | 0.49 | 0.52 | 0.55 | 0.85 | 0.84 | 0.79 | 0.73 | 0.74 |
| Board orientation relative to the ground | parallel | parallel | parallel | parallel | parallel | parallel | tilted | tilted |

[1] LP: low height and parallel, NP: normal height and parallel, NT: normal height and tilted.

### *B. Experimental Scenarios and Evaluation Metrics*

Two experimental scenarios are evaluated.

1) *With Position aiding*. During the first $60\ \text{s}$ of each dataset, the filter is updated with position measurements from the camera-based motion-tracking system. Thereafter, the filter relies solely on the IMU and magnetometer array. This scenario emulates the initial availability of GNSS or another positioning source. The root-mean-square errors (RMSEs) are computed over the samples with $t > 60\ \text{s}$, i.e., after position aiding ends, thereby quantifying each filter's ability to maintain accuracy without the position reference.

2) *Without position aiding*. Position measurements from any external positioning source are not used at any time, representing infrastructure-free autonomous indoor navigation. The RMSEs are computed over all valid samples with $t > 0\ \text{s}$.

The horizontal-position, vertical-position, velocity, and attitude RMSEs are reported. For a dataset with $N_d$ evaluation samples, the horizontal-position RMSE is

$$\text{RMSE}_{\text{hpos}} = \sqrt{\frac{1}{N_d}\sum_{k=1}^{N_d}\left\|\begin{bmatrix}1 & 0 & 0\\ 0 & 1 & 0\end{bmatrix}\left(\hat{\mathbf{p}}_k - \mathbf{p}_k\right)\right\|^2} \tag{86}$$

The vertical-position RMSE is

$$\mathrm{RMSE}_{\mathrm{vpos}} = \sqrt{\frac{1}{N_d}\sum_{k=1}^{N_d}\left\|\begin{bmatrix}0 & 0 & 1\end{bmatrix}\left(\hat{\mathbf{p}}_k - \mathbf{p}_k\right)\right\|^2} \tag{87}$$

The velocity RMSE is

$$\mathrm{RMSE}_{\mathrm{vel}} = \sqrt{\frac{1}{N_d}\sum_{k=1}^{N_d}\left\|\hat{\mathbf{v}}_k - \mathbf{v}_k\right\|^2} \tag{88}$$

The attitude RMSE is

$$\mathrm{RMSE}_{\mathrm{att}} = \sqrt{\frac{1}{N_d}\sum_{k=1}^{N_d}\alpha_k^2}, \quad \alpha_k = \left\|\log\left(\widehat{\mathbf{R}}_k \mathbf{R}_k^{\top}\right)\right\| \tag{89}$$

where $\log(\cdot)$ denotes the logarithmic map on $\mathrm{SO}(3)$, and $\alpha_k$ is the magnitude of the attitude error vector.

*C. Results With Position Aiding*

Taking LP-1 as an example, Fig. 3 shows the ground-truth and estimated horizontal trajectories under the five-sensor configuration. The trajectory estimated by the proposed RSE-KF-MIO tracks the ground truth more closely than that of MAINS. Fig. 4 shows the absolute horizontal-position, vertical-position, velocity, and attitude errors over time. The horizontal-position and attitude errors of RSE-KF-MIO remain below those of MAINS over most of the trajectory, whereas the vertical-position and velocity errors of the two filters are comparable. These results demonstrate the improved navigation accuracy of RSE-KF-MIO on LP-1. A quantitative comparison across all eight experiments follows in Table II.

Table II further summarizes the RMSEs of the two filters across all eight datasets. Relative to MAINS, the mean horizontal-position, velocity, and attitude RMSEs of RSE-KF-MIO are reduced by 33.3%, 36.8%, and 46.0% with the five-sensor configuration. The vertical position behaves differently, and the mean RMSEs of the two filters are comparable (2.94 m vs. 3.01 m). With the 30-sensor configuration, the corresponding reductions in the horizontal-position, velocity, and attitude RMSEs are 31.6%, 35.3%, and 28.2%, and the vertical-position RMSE is reduced by 16.8%. These results demonstrate that RSE-KF-MIO achieves better overall navigation accuracy than MAINS across the eight datasets under both sensor configurations.

The improvement of RSE-KF-MIO over MAINS varies across the eight datasets. Given the limited number of datasets, the variation cannot yet be associated with differences in the experimental settings, such as the inclination angle and height of the sensor board. However, since the two methods use the same nominal state propagation and basic filter parameters, it can be inferred that this variation arises from the different ways the two filters respond to each trajectory's motion and magnetic-field conditions. Specifically, in regions where the magnetic field varies weakly in space, the magnetometer measurements may provide limited pose information, and both filters may degrade toward pure SINS drift, resulting in a small difference between the two. By contrast, when the magnetic field provides informative pose information while the SINS nominal state deviates considerably from the true state, the advantage of the proposed RSE-KF-MIO over MAINS tends to become more pronounced, since its right-invariant error depends only weakly on the nominal state, leading to slower error accumulation.

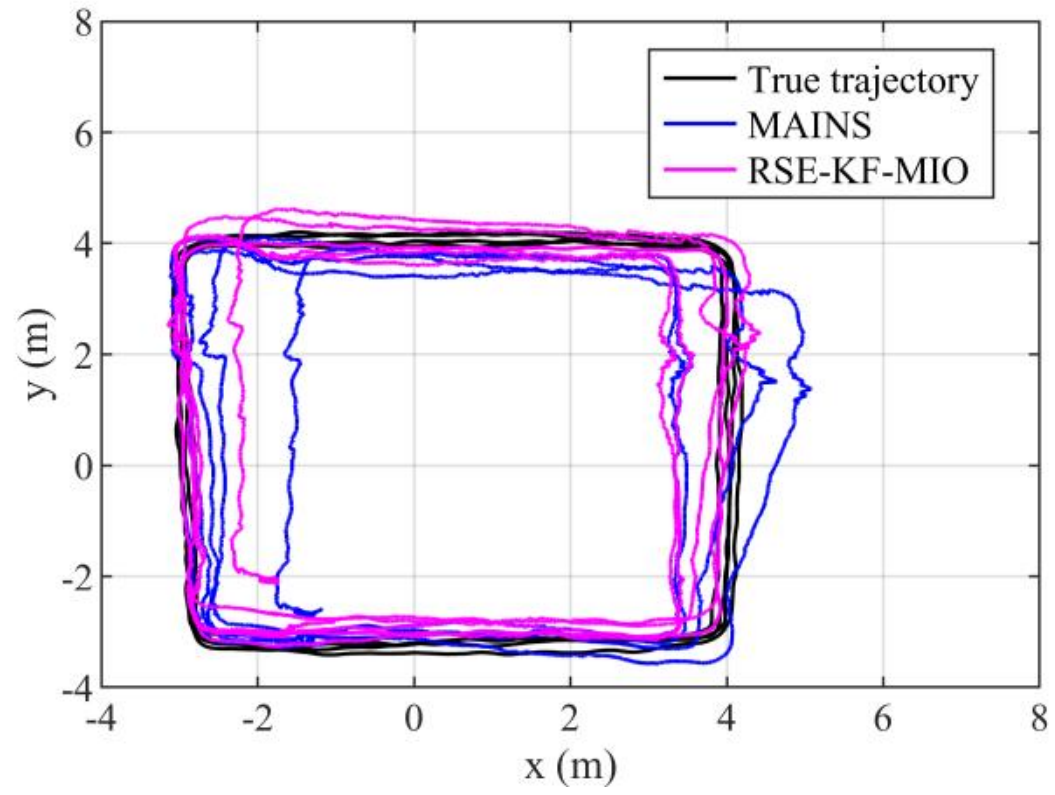

**Fig. 3.** Horizontal trajectories of MAINS and RSE-KF-MIO versus the ground truth on LP-1 under the five-sensor configuration with position aiding.

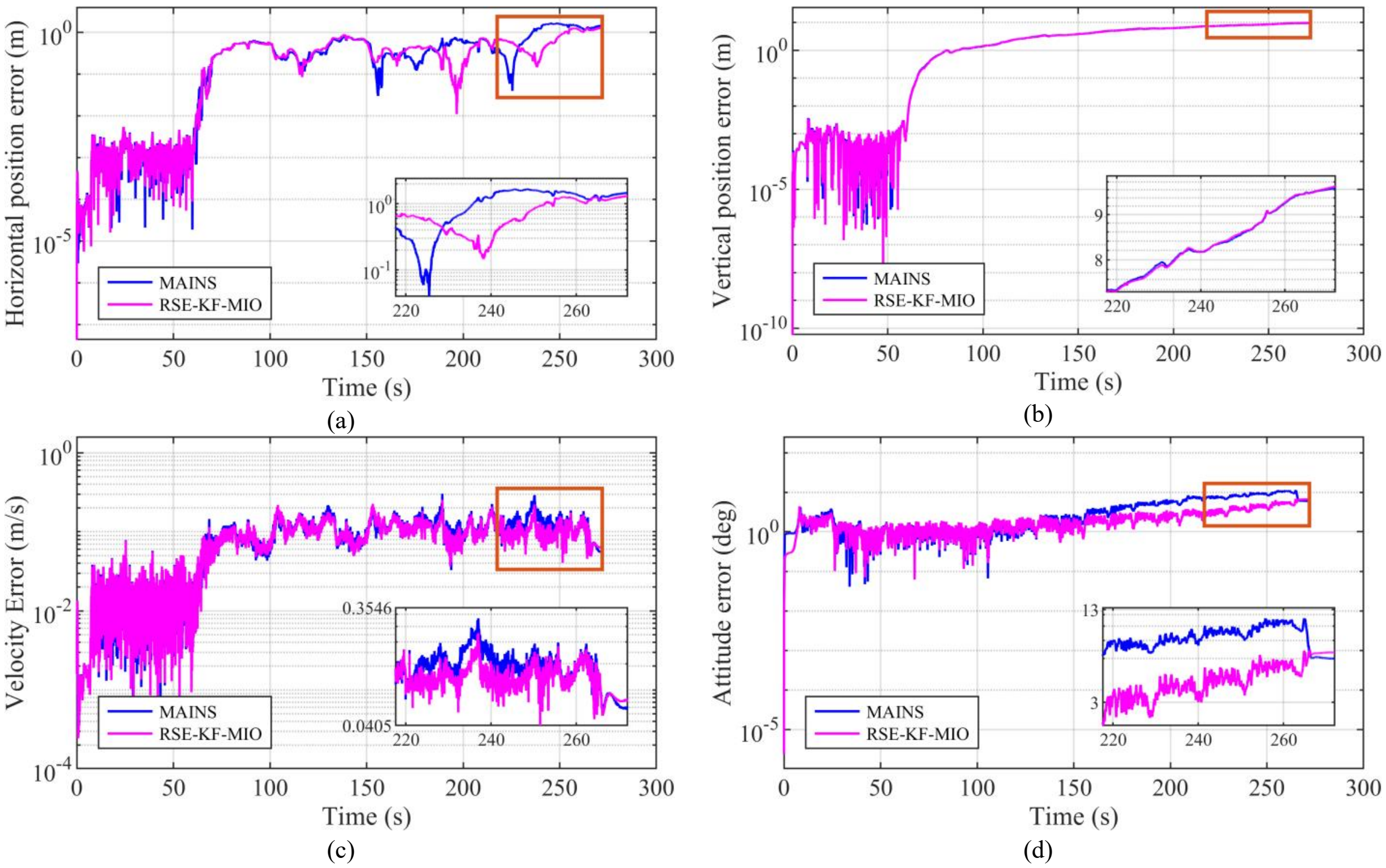

**Fig. 4.** Absolute errors of horizontal-position (a), vertical-position (b), velocity (c), and attitude (d) over time on LP-1 under the five-sensor configuration with position aiding.

TABLE II
RMSE Comparison With Position Aiding

| Dataset | LP-1 | LP-2 | LP-3 | NP-1 | NP-2 | NP-3 | NT-1 | NT-2 | MEAN[2] |
|---|---|---|---|---|---|---|---|---|---|
| Trajectory length (m)[1] | 114.14 | 139.87 | 162.3 | 93.68 | 89.47 | 86.43 | 123.01 | 87.26 | |
| Trajectory duration (s)[1] | 212 | 226 | 272 | 117 | 105 | 94 | 125 | 91 | |
| MAINS with the **5**-sensor configuration | | | | | | | | | |
| Horizontal position RMSE (m) | 0.72 | 1.42 | 1.78 | 1.96 | 1.71 | 1.95 | 2.67 | 0.75 | 1.62 |

| Dataset | LP-1 | LP-2 | LP-3 | NP-1 | NP-2 | NP-3 | NT-1 | NT-2 | MEAN[2] |
|---|---|---|---|---|---|---|---|---|---|
| Trajectory length (m)[1] | 114.14 | 139.87 | 162.3 | 93.68 | 89.47 | 86.43 | 123.01 | 87.26 | |
| Trajectory duration (s)[1] | 212 | 226 | 272 | 117 | 105 | 94 | 125 | 91 | |
| Vertical position RMSE (m) | 5.60 | 6.36 | 6.91 | 1.46 | 0.55 | 1.21 | 1.28 | 0.70 | 3.01 |
| Velocity RMSE (m/s) | 0.12 | 0.17 | 0.14 | 0.24 | 0.17 | 0.29 | 0.22 | 0.20 | 0.19 |
| Attitude RMSE (deg) | 5.21 | 10.77 | 6.71 | 2.71 | 4.64 | 3.35 | 1.82 | 6.38 | 5.20 |
| RSE-KF-MIO with the **5**-sensor configuration | | | | | | | | | |
| Horizontal position RMSE (m) | 0.57 | 1.20 | 1.67 | 1.33 | 1.40 | 1.01 | 0.74 | 0.72 | **1.08** |
| Vertical position RMSE (m) | 5.59 | 6.34 | 6.88 | 1.25 | 0.24 | 1.08 | 1.40 | 0.72 | **2.94** |
| Velocity RMSE (m/s) | 0.11 | 0.15 | 0.11 | 0.12 | 0.12 | 0.13 | 0.12 | 0.11 | **0.12** |
| Attitude RMSE (deg) | 2.80 | 8.27 | 1.85 | 2.67 | 1.76 | 1.30 | 2.22 | 1.59 | **2.81** |
| MAINS with the **30**-sensor configuration | | | | | | | | | |
| Horizontal position RMSE (m) | 0.71 | 0.90 | 1.29 | 2.79 | 1.27 | 2.40 | 1.05 | 1.96 | 1.55 |
| Vertical position RMSE (m) | 0.48 | 0.60 | 0.73 | 2.15 | 0.32 | 2.18 | 1.89 | 0.24 | 1.07 |
| Velocity RMSE (m/s) | 0.11 | 0.11 | 0.12 | 0.20 | 0.24 | 0.24 | 0.19 | 0.17 | 0.17 |
| Attitude RMSE (deg) | 4.97 | 3.84 | 8.28 | 3.09 | 4.16 | 1.16 | 1.57 | 2.16 | 3.65 |
| RSE-KF-MIO with the **30**-sensor configuration | | | | | | | | | |
| Horizontal position RMSE (m) | 0.88 | 0.90 | 1.35 | 1.87 | 1.42 | 1.11 | 0.35 | 0.61 | **1.06** |
| Vertical position RMSE (m) | 0.53 | 0.61 | 0.94 | 1.45 | 0.89 | 1.34 | 1.11 | 0.20 | **0.89** |
| Velocity RMSE (m/s) | 0.09 | 0.15 | 0.11 | 0.12 | 0.10 | 0.12 | 0.09 | 0.11 | **0.11** |
| Attitude RMSE (deg) | 2.04 | 7.53 | 2.44 | 2.51 | 1.77 | 0.84 | 2.28 | 1.53 | **2.62** |

[1] Statistics exclude the first 60-s position-aiding window; [2] The MEAN column gives the arithmetic mean of the RMSEs over the eight datasets.

*D. Results Without Position Aiding*

Taking LP-1 as an example, Fig. 5 shows the ground-truth and estimated horizontal trajectories of MAINS and RSE-KF-MIO under the five-sensor configuration without position aiding. Without position aiding, the trajectories estimated by both filters deviate noticeably from the ground truth. Fig. 6 further shows the absolute horizontal-position, vertical-position, velocity, and attitude errors over time. It can be observed that the proposed RSE-KF-MIO exhibits lower horizontal-position, velocity, and attitude errors than MAINS over most of the trajectory. Compared with the position-aiding scenario, the error gap between the two filters is further widened, indicating that the navigation performance of MAINS degrades more severely without position aiding.

Table III summarizes the errors of the two filters across all eight datasets without position aiding. Overall, both filters exhibit larger navigation errors than in the position-aiding scenario. Without the first 60 s of position aiding, the initial errors and subsequent drift are less constrained by external measurements, inevitably degrading filter performance.

The two filters exhibit markedly different performance. Relative to MAINS, RSE-KF-MIO reduces the horizontal-position, velocity, and attitude RMSEs by 55.0%, 57.4%, and 62.5% with the five-sensor configuration, while the reduction in the vertical-position RMSE is 6.0%. With the 30-sensor configuration, the corresponding reductions in the horizontal-position, velocity, and attitude RMSEs are 44.1%, 50.0%, and 54.0%, and the reduction in the vertical-position RMSE is 14.7%. The performance difference becomes more pronounced than in the position-aiding scenario because the nominal-state estimates are more prone to inaccuracy when the magnetic field is the only external information source. Specifically, the attitude-error equation of MAINS explicitly involves the current estimated angular velocity, whereas, owing to the right-invariant error modeling, that of RSE-KF-MIO depends on the nominal state only through the gyroscope-bias term. When the nominal state deviates, the attitude error of MAINS may grow, and the inaccurate attitude estimate is subsequently propagated into the velocity and position recursions, causing further error accumulation. This effect is particularly evident on NT-2 with the five-sensor configuration. As shown in Fig. 7, MAINS exhibits substantial attitude errors in the initial stage of the motion, and its trajectory deviates severely from the ground truth. In contrast, the absolute attitude error of RSE-KF-MIO remains mostly below 10° , and its estimated trajectory stays relatively close to the ground truth. Consistent with the position-aiding scenario, RSE-KF-MIO again achieves only a limited reduction in the vertical-position RMSE relative to MAINS.

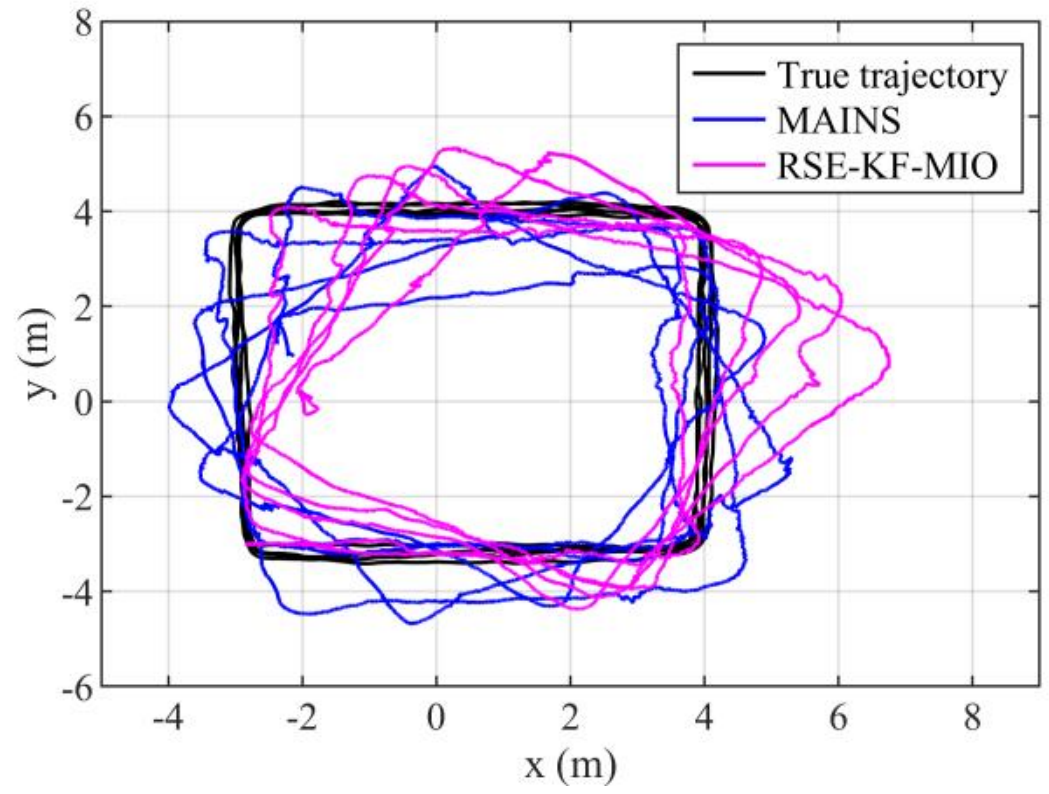

**Fig. 5.** Horizontal trajectories of MAINS and RSE-KF-MIO versus the ground truth on LP-1 under the five-sensor configuration without position aiding.

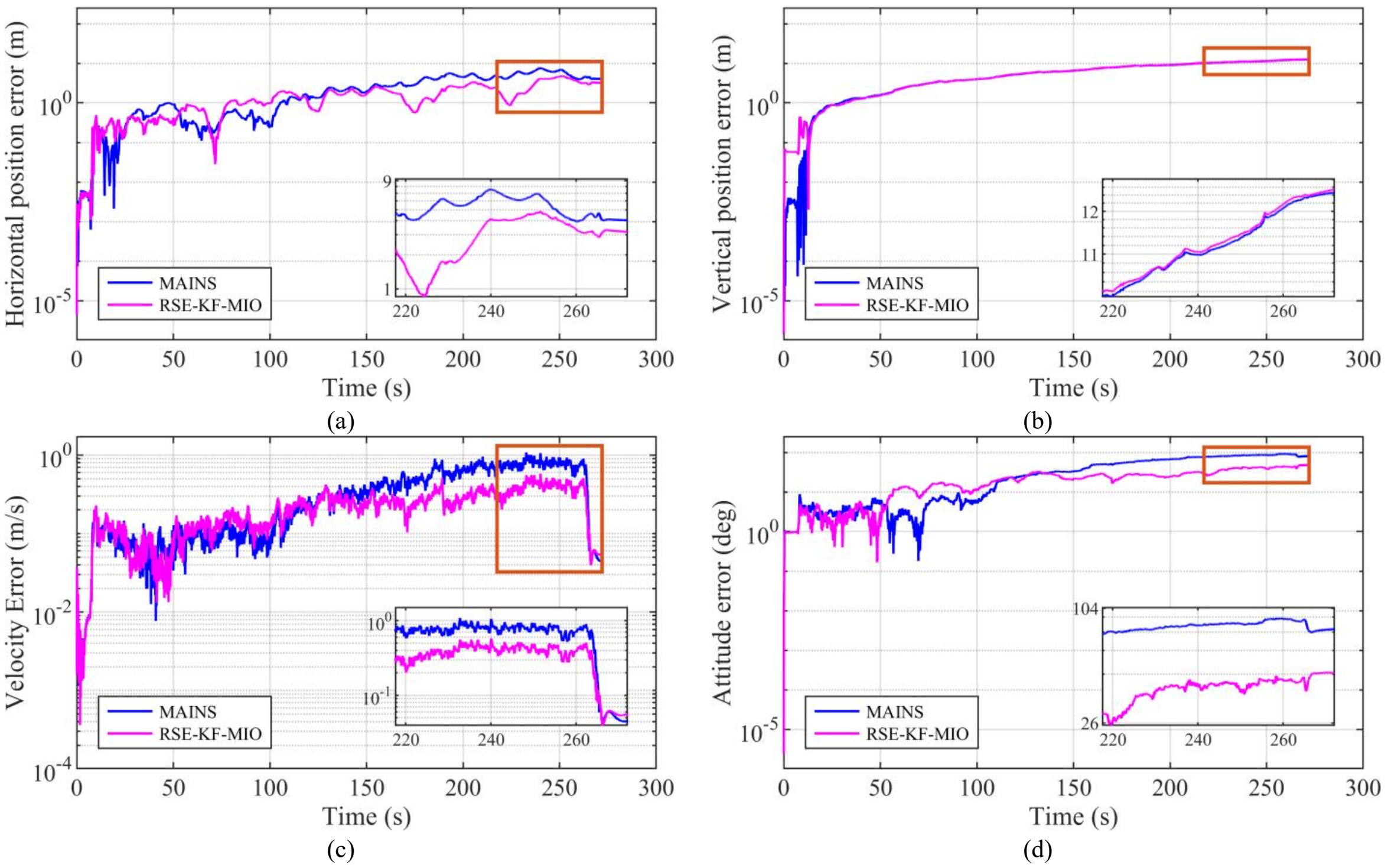

**Fig. 6.** Absolute errors of horizontal-position (a), vertical-position (b), velocity (c), and attitude (d) over time on LP-1 under the five-sensor configuration without position aiding.

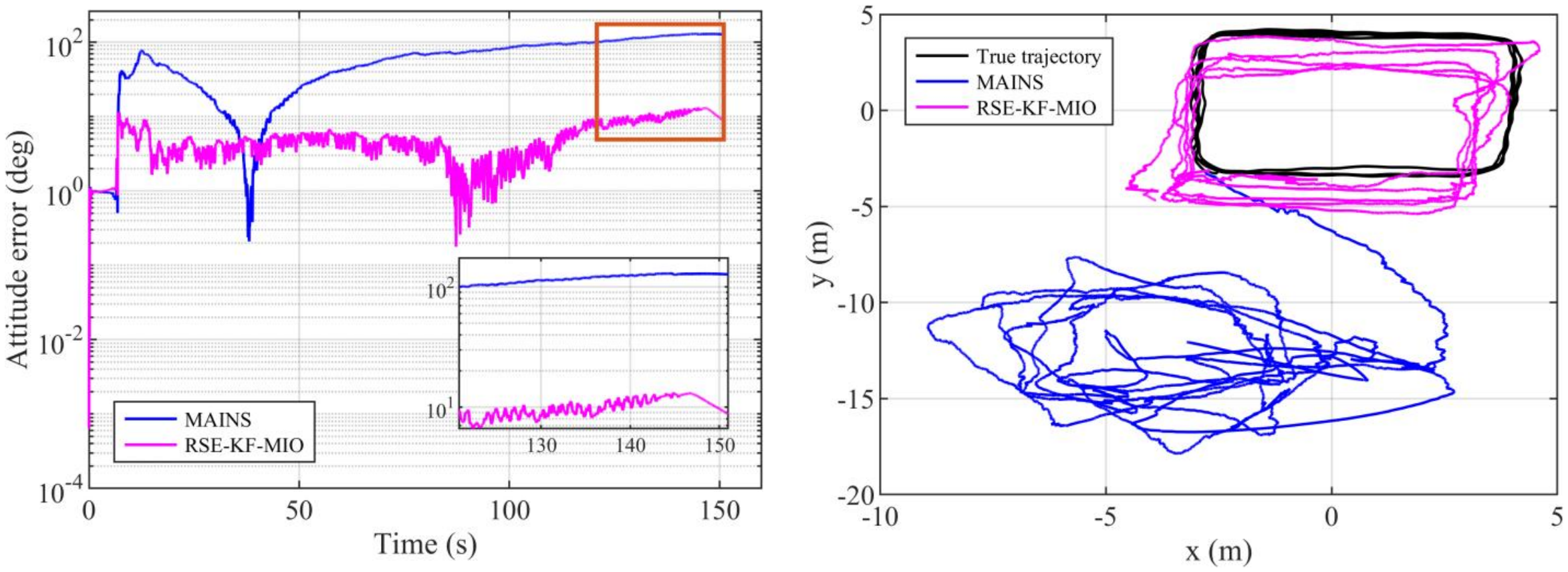

(a) (b)

**Fig. 7.** Absolute attitude error (a) and horizontal trajectory (b) of MAINS and RSE-KF-MIO on NT-2 under the five-sensor configuration without position aiding.

TABLE III
RMSE COMPARISON WITHOUT POSITION AIDING

| **Dataset** | **LP-1** | **LP-2** | **LP-3** | **NP-1** | **NP-2** | **NP-3** | **NT-1** | **NT-2** | **MEAN**[2] |
|---|---|---|---|---|---|---|---|---|---|
| Trajectory length (m) | 138.72 | 167.07 | 194.41 | 136.23 | 134.66 | 137.76 | 164.62 | 137.87 | |
| Trajectory duration (s)[1] | 272 | 286 | 332 | 177 | 165 | 154 | 185 | 151 | |
| MAINS with the **5**-sensor configuration | | | | | | | | | |
| Horizontal position RMSE (m) | 3.23 | 3.69 | 3.37 | 3.15 | 2.15 | 3.48 | 2.97 | 13.73 | 4.47 |
| Vertical position RMSE (m) | 7.12 | 7.85 | 8.54 | 2.07 | 0.76 | 1.18 | 1.14 | 4.68 | 4.17 |
| Velocity RMSE (m/s) | 0.45 | 0.55 | 0.48 | 0.53 | 0.30 | 0.81 | 0.22 | 1.51 | 0.61 |
| Attitude RMSE (deg) | 49.29 | 52.95 | 46.42 | 42.89 | 17.13 | 53.32 | 6.03 | 74.86 | 42.86 |
| RSE-KF-MIO with the **5**-sensor configuration | | | | | | | | | |
| Horizontal position RMSE (m) | 1.96 | 2.67 | 2.95 | 1.32 | 2.62 | 1.34 | 1.72 | 1.47 | **2.01** |
| Vertical position RMSE (m) | 7.16 | 7.89 | 8.58 | 2.11 | 1.04 | 2.30 | 0.89 | 1.42 | **3.92** |
| Velocity RMSE (m/s) | 0.25 | 0.36 | 0.35 | 0.26 | 0.24 | 0.22 | 0.14 | 0.29 | **0.26** |
| Attitude RMSE (deg) | 25.05 | 32.66 | 30.19 | 13.50 | 6.33 | 10.47 | 4.35 | 6.01 | **16.07** |
| MAINS with the **30**-sensor configuration | | | | | | | | | |
| Horizontal position RMSE (m) | 1.65 | 2.20 | 4.05 | 5.09 | 3.88 | 4.29 | 1.85 | 2.17 | 3.15 |
| Vertical position RMSE (m) | 0.83 | 1.05 | 1.48 | 3.19 | 0.89 | 2.53 | 1.97 | 0.66 | 1.57 |
| Velocity RMSE (m/s) | 0.19 | 0.30 | 0.56 | 0.78 | 0.68 | 0.77 | 0.35 | 0.42 | 0.50 |

| Dataset | LP-1 | LP-2 | LP-3 | NP-1 | NP-2 | NP-3 | NT-1 | NT-2 | MEAN[2] |
|---|---|---|---|---|---|---|---|---|---|
| Trajectory length (m) | 138.72 | 167.07 | 194.41 | 136.23 | 134.66 | 137.76 | 164.62 | 137.87 | |
| Trajectory duration (s)[1] | 272 | 286 | 332 | 177 | 165 | 154 | 185 | 151 | |
| Attitude RMSE (deg) | 19.02 | 31.14 | 63.66 | 63.11 | 44.68 | 53.97 | 16.92 | 21.60 | 39.26 |
| RSE-KF-MIO with the **30**-sensor configuration | | | | | | | | | |
| Horizontal position RMSE (m) | 2.46 | 3.06 | 2.29 | 2.13 | 1.44 | 1.30 | 0.45 | 0.95 | **1.76** |
| Vertical position RMSE (m) | 0.90 | 1.25 | 1.66 | 2.66 | 1.33 | 2.17 | 0.40 | 0.34 | **1.34** |
| Velocity RMSE (m/s) | 0.32 | 0.45 | 0.36 | 0.17 | 0.22 | 0.19 | 0.10 | 0.17 | **0.25** |
| Attitude RMSE (deg) | 33.75 | 41.25 | 31.37 | 7.23 | 8.90 | 8.27 | 4.19 | 9.59 | **18.07** |

[1] Statistics include the entire time window. [2] The MEAN column gives the arithmetic mean of the RMSEs over the eight datasets.

*E. Sensor-Count Sweep Experiment*

To investigate the effect of the sensor count on navigation performance, a sweep experiment was conducted over the number of active sensors. Fig. 8 shows the configurations used in the sweep experiment, where the active sensors were selected to maintain a spatially dispersed layout over the board. Fig. 9 and 10 show the mean RMSEs of the four metrics as functions of the sensor count, with and without position aiding, respectively. The trends are broadly consistent between the two scenarios. For both filters, the mean RMSEs first decrease as the sensor count increases and then level off. Most metrics stabilize at around seven sensors, although a few continue to fluctuate at larger counts. Except for the horizontal-position RMSE with three sensors in the position-aiding scenario, the proposed RSE-KF-MIO achieves lower mean horizontal-position, velocity, and attitude RMSEs than MAINS in all the tested configurations, although the improvement in the vertical-position RMSE remains limited. These results demonstrate a stable advantage of the proposed method across the sensor configurations in both scenarios.

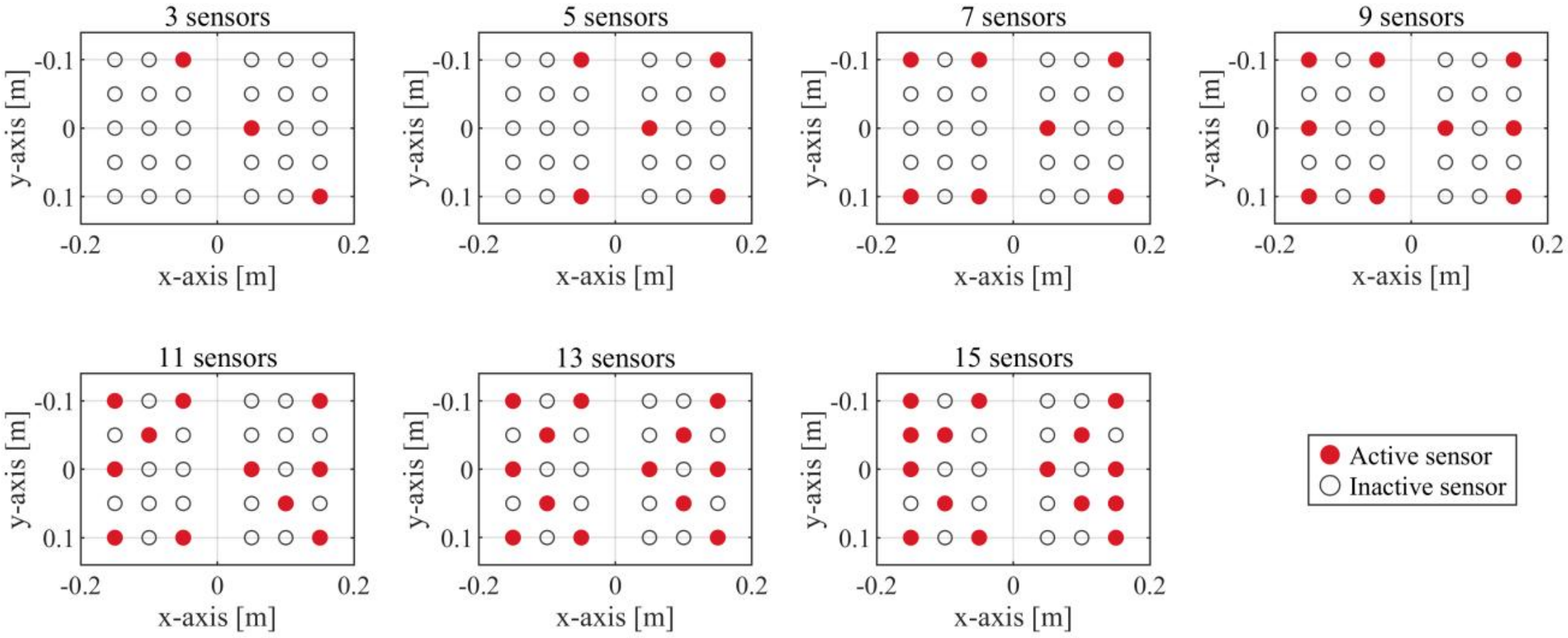

**Fig. 8.** Sensor configurations used in the sensor-count sweep experiment.

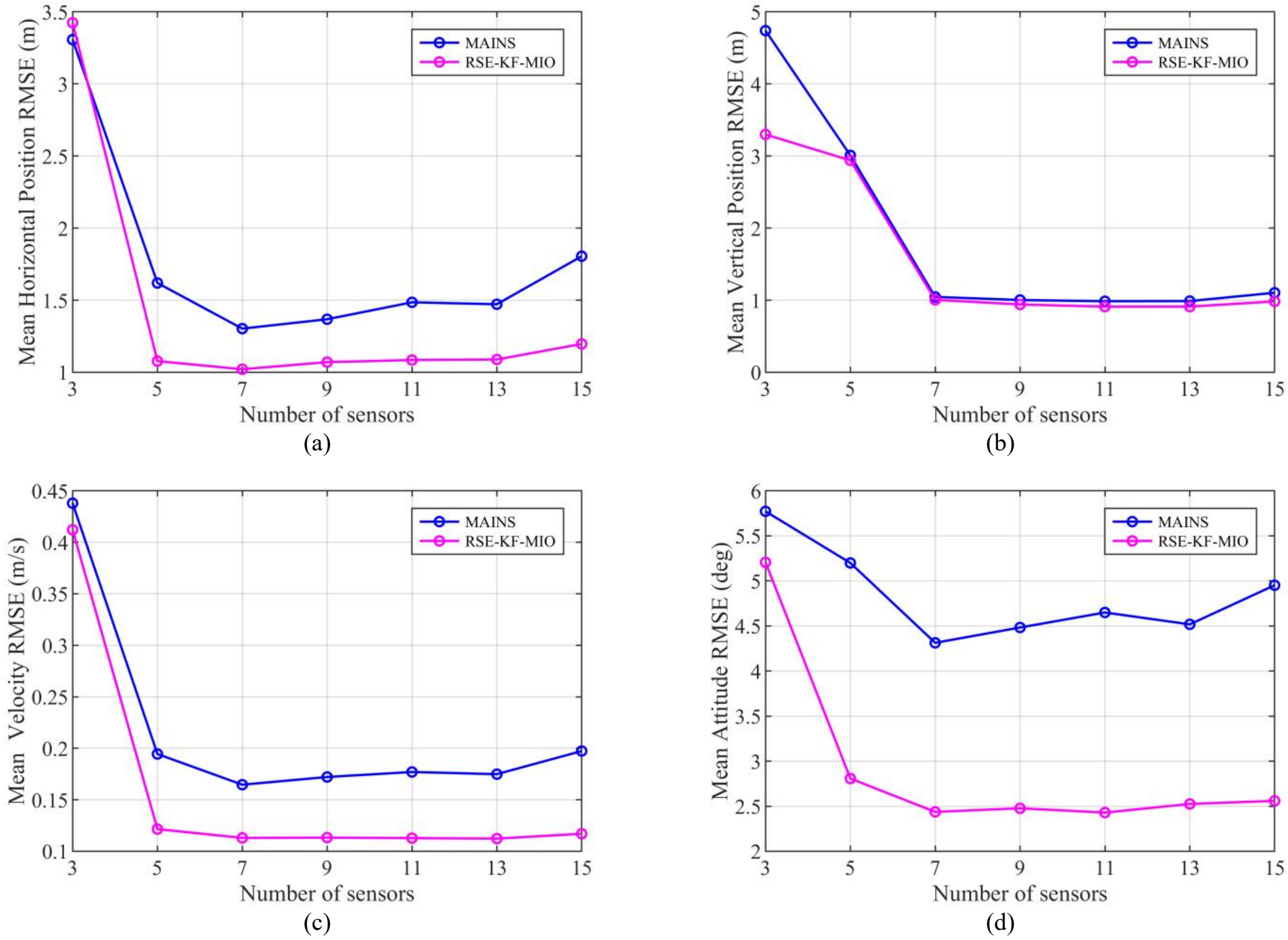

**Fig. 9.** Mean RMSEs of the two filters as functions of the sensor count with position aiding. (a) Horizontal position. (b) Vertical position. (c) Velocity. (d) Attitude.

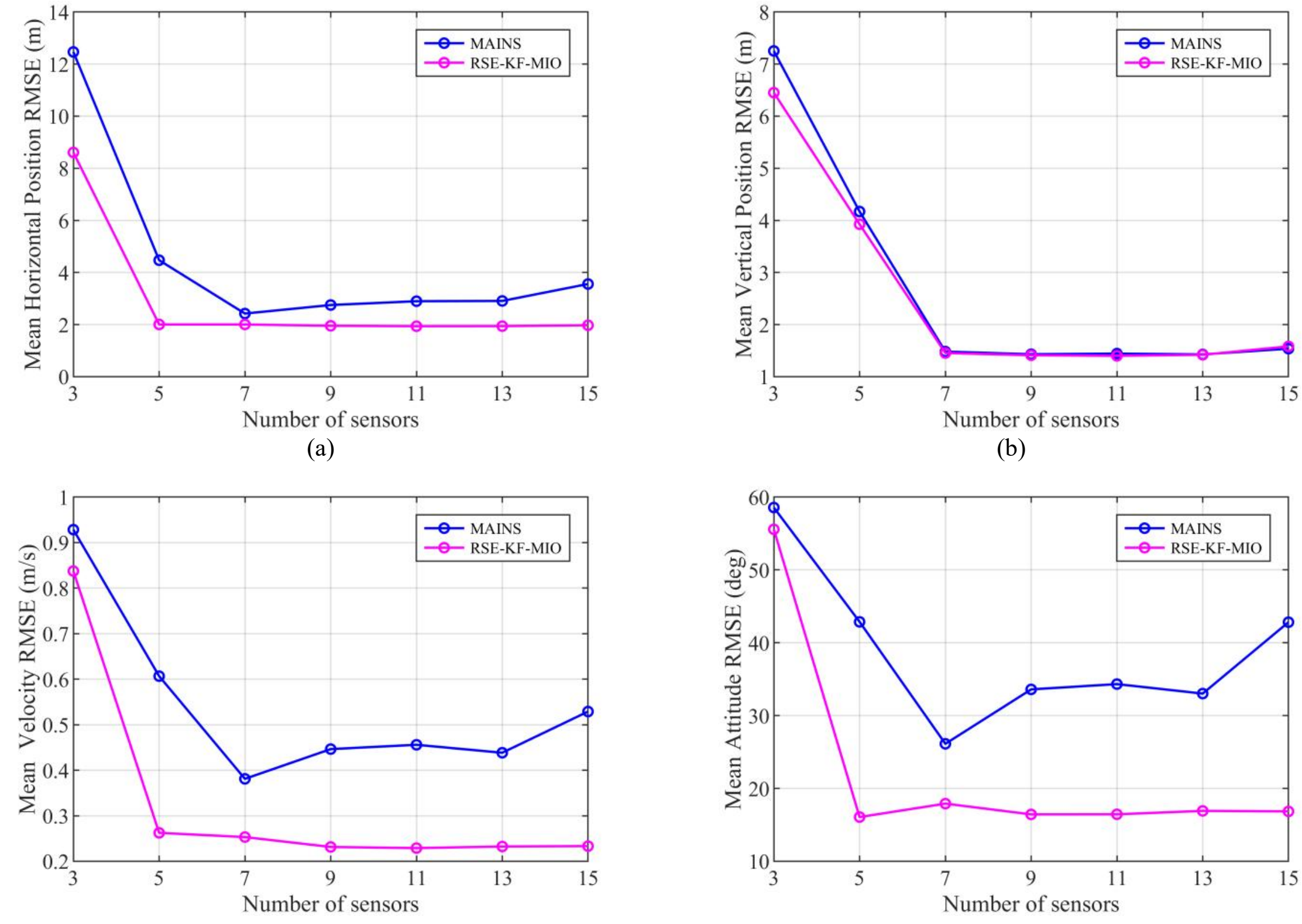

(c) (d)

**Fig. 10.** Mean RMSEs of the two filters as functions of the sensor count without position aiding. (a) Horizontal position. (b) Vertical position. (c) Velocity. (d) Attitude.

### *F. Polynomial-Model Coefficient Bias Ablation Experiment*

A polynomial-model coefficient bias was introduced to compensate for the propagation error of the polynomial-model coefficients. Its effectiveness is evaluated through an ablation experiment with two settings, namely RSE-KF-MIO without the polynomial-model coefficient bias (RSE-KF-MIO-NoBias) and MAINS with the polynomial-model coefficient bias added (MAINS-WithBias). This cross-comparison evaluates the contribution of the bias term in combination with the right-invariant error modeling. Table IV lists the results with five-sensor configuration.

Removing the bias state from RSE-KF-MIO increases the mean horizontal-position RMSE from 1.08 m to 1.52 m with position aiding and from 2.01 m to 2.54 m without position aiding. The same removal also increases the mean velocity and attitude RMSEs in both scenarios. Conversely, adding the bias to MAINS reduces the mean horizontal-position RMSE from 1.62 m to 1.13 m with position aiding and from 4.47 m to 3.99 m without position aiding, . The velocity and attitude RMSEs also decrease in both scenarios, with the only exception of a slight increase in the vertical-position RMSE without position aiding (from 4.17 m to 4.41 m). These results indicate that the polynomial-model coefficient bias effectively compensates for the drift arising during coefficient propagation.

Nevertheless, MAINS-WithBias still exhibits higher mean horizontal-position, velocity, and attitude RMSEs than RSE-KF-MIO, and the gap becomes much larger without position aiding, e.g., 3.99 m versus 2.01 m in the horizontal position. This indicates that the coefficient bias alone cannot compensate for the linearization error of the conventional ESKF. The performance gain of RSE-KF-MIO therefore stems from two complementary sources, namely the coefficient bias that absorbs the model mismatch accumulated during coefficient propagation and, more importantly, the right-invariant error modeling whose weaker dependence on the nominal state becomes the dominant contributor when position aiding is unavailable.

TABLE IV
RESULTS OF THE POLYNOMIAL-MODEL COEFFICIENT BIAS ABLATION EXPERIMENT WITH FIVE-SENSOR CONFIGURATION

| | With Position aiding / without position aiding | | | |
|---|---|---|---|---|
| | Mean horizontal position RMSE (m) | Mean vertical position RMSE (m) | Mean velocity RMSE (m/s) | Mean attitude RMSE (deg) |
| RSE-KF-MIO | 1.08 /2.01 | 2.94 /3.92 | 0.12 /0.26 | 2.81 /16.07 |
| RSE-KF-MIO-NoBias | 1.52 /2.54 | 3.07 /3.93 | 0.18 /0.36 | 3.31 /25.37 |
| | | | | |
| MAINS | 1.62 /4.47 | 3.01 /4.17 | 0.19 /0.61 | 5.20 /42.86 |
| MAINS-WithBias | 1.13 /3.99 | 2.94 /4.41 | 0.13 /0.53 | 4.33 /35.06 |

## VI. CONCLUSION

This article presented RSE-KF-MIO, which combines invariant error-state Kalman filtering on Lie group $SE_2(3)$ with a scaled magnetic-field model coefficient bias for map-free indoor navigation. Under identical nominal recursions, magnetic-field models, basic filter parameters, and input data, RSE-KF-MIO achieved better overall horizontal position, velocity, and attitude accuracy than the ESKF-based MAINS estimator across the eight datasets in both aiding scenarios. In particular, the larger gains observed without position aiding are consistent with the reduced dependence of right-invariant error propagation on the nominal state. The sensor-count sweep experiment showed that RSE-KF-MIO outperformed MAINS in most sensor-count configurations. Most accuracy metrics stabilized at around seven magnetometers, beyond which additional sensors brought little further improvement, and MAINS even exhibited slight degradation at larger sensor counts, indicating that the conventional ESKF is more susceptible to the disturbances introduced by the additional measurements. Furthermore, the ablation study confirms that the improvement stems from two complementary sources, namely the coefficient-bias compensation, which accounts for a substantial part of the accuracy gain, particularly with position aiding, and the invariant error modeling, which is the dominant contributor to the larger gains observed without position aiding.

Despite these improvements, the two filters achieved comparable vertical position accuracy, indicating that the proposed estimator modifications had a limited effect on vertical estimation. This limitation may be partly associated with the restricted vertical spatial information provided by the planar arrays and tested trajectories. Future work will therefore develop a quantitative observability assessment method for magnetic-field navigation, based on which the adaptability of magnetometer-array configurations and spatial layouts to different magnetic-field environments will be investigated.